\documentclass{aa}  
\usepackage{silence}
\usepackage{graphicx}
\usepackage{txfonts}
\usepackage[dvipsnames]{xcolor}

\usepackage{lscape}
\usepackage{longtable}
\usepackage[colorlinks=true,     linkcolor=blue, citecolor=blue, filecolor=blue, urlcolor=blue]{hyperref}
\usepackage{xparse}
\usepackage[normalem]{ulem}
\usepackage{multirow}

\usepackage{listings}

\definecolor{codegreen}{rgb}{0,0.6,0}
\definecolor{codegray}{rgb}{0.5,0.5,0.5}
\definecolor{codepurple}{rgb}{0.58,0,0.82}
\definecolor{codebkg}{rgb}{0.95,0.95,0.92}

\lstdefinestyle{listingstyle}{
    backgroundcolor=\color{codebkg},
    commentstyle=\color{codegreen},
    keywordstyle=\color{blue},
    numberstyle=\tiny\color{codegray},
    stringstyle=\color{codepurple},
    basicstyle=\ttfamily\footnotesize,
    breakatwhitespace=false,
    showspaces=false,
    showstringspaces=false,
    showtabs=false,
    tabsize=2
}
\begin{document} 

   \title{CAVITY: Calar Alto Void Integral-field Treasury surveY: II. Second public data release}
   \titlerunning{CAVITY DR2}

   \author{B. Bidaran\inst{\ref{ugr1}}
   \and A. Jiménez\inst{\ref{ugr1}}
\and R. García-Benito\inst{\ref{iaa}}
\and M. Argudo-Fernández\inst{\ref{ugr1},\ref{ugr2}}
\and M. Alcázar-Laynez\inst{\ref{ugr1}}
\and S.B. De Daniloff\inst{\ref{iram},\ref{ugr1}}
\and S. Duarte Puertas\inst{\ref{ugr1},\ref{ugr2},\ref{laval}}
 \and D. Espada\inst{\ref{ugr1},\ref{ugr2}}
\and E. Florido\inst{\ref{ugr1},\ref{ugr2}}
\and Y. K.  González-Koda\inst{\ref{ugr1}}
\and I. Pérez\inst{\ref{ugr1},\ref{ugr2}}
\and T. Ruiz-Lara\inst{\ref{ugr1},\ref{ugr2}}
\and L. Sánchez-Menguiano\inst{\ref{ugr1},\ref{ugr2}}
\and G. Torres-Ríos\inst{\ref{ugr1}}
\and P. Vásquez-Bustos\inst{\ref{ugr1}}
\and S. Verley\inst{\ref{ugr1},\ref{ugr2}}
 \and A. Zurita\inst{\ref{ugr1},\ref{ugr2}}
 \and M. A. Canossa-Gosteinski\inst{\ref{kapteyn}}
\and A. M.~Conrado\inst{\ref{iaa}}
 \and H. M. Courtois\inst{\ref{cbl}}
 \and I. del Moral-Castro\inst{\ref{chile}}
\and J. Falc\'on-Barroso\inst{\ref{iac},\ref{ull}}
 \and A. Ferr\'e-Mateu\inst{\ref{iac},\ref{ull}}
\and M. Hernández-Sánchez\inst{\ref{UV}}
\and A. Lugo-Aranda\inst{\ref{unam}}
 \and S. Subramanian\inst{\ref{iia},\ref{AIP}}
\and P. Villalba-González\inst{\ref{columbia}}
\and L. Galbany\inst{\ref{ice},\ref{ieec}}
\and R. M.~González Delgado\inst{\ref{iaa}}
\and R. F. Peletier\inst{\ref{kapteyn}}
\and S. F. Sánchez\inst{\ref{ull},\ref{unam}}
\and R. van de Weygaert\inst{\ref{kapteyn}}
\and A. Fernández-Martín\inst{\ref{caha}}}

   \institute{
Dpto. de F\'{\i}sica Te\'orica y del Cosmos, Facultad de Ciencias (Edificio Mecenas), Universidad de Granada, E-18071, Granada, Spain\label{ugr1}
\and Instituto de Astrof\'isica de Andaluc\'ia - CSIC, Glorieta de la Astronomía s/n, E-18008 Granada, Spain\label{iaa} 
\and Instituto Carlos I de F\'\i sica Te\'orica y Computacional, Universidad de Granada, E-18071, Granada, Spain\label{ugr2}
\and Institut de Radioastonomie Millim\'etrique (IRAM), Av. Divina Pastora 7, N\'ucleo Central E-18012, Granada, Spain\label{iram}
\and D\'epartement de Physique, de G\'enie Physique et d’Optique, Universit\'e Laval, and Centre de Recherche en Astrophysique du Qu\'ebec (CRAQ), Québec, QC, G1V 0A6, Canada\label{laval}
\and Kapteyn Astronomical Institute, University of Groningen, PO Box 800, 9700 AV Groningen, The Netherlands\label{kapteyn}
\and Universit\'e Claude Bernard Lyon 1, IUF, IP2I Lyon, 4 rue Enrico Fermi, 69622 Villeurbanne, France\label{cbl}
\and Instituto de Astrof\'isica, Facultad de F\'isica, Pontificia Universidad Cat\'olica de Chile, Campus San Joaqu\'in, Av. Vicu\~na Mackenna 4860, Macul, Santiago, Chile, 7820436\label{chile}
\and Instituto de Astrof\'isica de Canarias, c/V\'ia L\'actea s/n, E-38205, La Laguna, Tenerife, Spain\label{iac}
\and Departamento de Astrof\'isica, Universidad de La Laguna, E-38206, La Laguna, Tenerife, Spain\label{ull}
\and Departament d’Astronomia i Astrof\'isica, Universitat de València, E-46100 Burjassot (València), Spain\label{UV}
\and Instituto de Astronomía, Universidad Nacional Autonóma de México, A.P. 70-264, 04510 Ciudad de México, México\label{unam}
\and Indian Institute of Astrophysics, Koramangala II Block, Bangalore-560034, India\label{iia}
\and Leibniz-Institut für Astrophysik Potsdam (AIP), An der Sternwarte 16, D-14482 Potsdam, Germany \label{AIP}
\and Department of Physics and Astronomy, University of British Columbia, Vancouver, BC V6T 1Z1, Canada\label{columbia}
\and Institute of Space Sciences (ICE-CSIC), Campus UAB, Carrer de Can Magrans, s/n, E-08193 Barcelona, Spain\label{ice}
\and Institut d'Estudis Espacials de Catalunya (IEEC), E-08860 Castelldefels (Barcelona), Spain\label{ieec}
\and Centro Astronómico Hispano en Andalucía, Observatorio de Calar Alto, Sierra de los Filabres, E-04550 Gérgal, Almería, Spain\label{caha}
    }
    
   \date{Received Month Day, Year; accepted Month Day, Year}

  \abstract
   {{Context.} Studying galaxy evolution under the unique environmental conditions of cosmic voids provides an opportunity to disentangle the role of the large-scale environment in shaping mass assembly (both baryonic and dark matter), regulating galaxy physical properties, and driving the transformation from star-forming to quiescent systems.
   
   {Aims.} We present the second public data release (DR2) of the Calar Alto Void Integral-field Treasury Survey (CAVITY), an ongoing legacy programme designed to study void galaxies (VGs) across a range of nearby ($0.005 \leq z \leq 0.050$) voids with different sizes and dynamical stages, providing 200 science-grade optical integral-field spectroscopy (IFS) data cubes. {By doubling the number of galaxies relative to DR1 while maintaining the same selection criteria, DR2 significantly increases the dataset’s statistical power, enabling more robust characterization of galaxy properties and their correlations with environment.}
   
   {Methods.} The observations were obtained with the PMAS/PPAK spectrograph on the 3.5-m Calar Alto telescope using the V500 configuration, covering the optical spectral range of 3745-7500 \AA\ at a resolution of 6 \AA\ (FWHM). 
   
   {Results.} The DR2 VG sample spans a broad range of stellar masses, morphologies, colours, and gas ionisation properties. We describe the sample selection, observing strategy, data reduction pipeline, quality-control procedures, and the public access to the CAVITY datasets and associated ancillary data on the survey's database. In addition, we present a characterisation of the large- and local-scale environments of CAVITY galaxies, showing that the 15 surveyed voids encompass a wide variety of sizes, galaxy richness, galaxy populations, and local structures, including galaxy groups.
   
   {Conclusions.} This release comprises 200 IFS data cubes, publicly available together with the master catalogues at the survey’s dedicated webpage: \url{https://cavity.caha.es/data/dr2/}.}


   \maketitle

\defcitealias{2024A&A...691A.161G}{DR1} 	

\section{Introduction}\label{Introduction}

The uneven distribution of galaxies on scales larger than 10\,Mpc traces the underlying inhomogeneous dark matter distribution that gives rise to the cosmic web \citep{1996Natur.380..603B,2014MNRAS.441.2923C}. In this web-like structure, most of the matter is concentrated within elongated filaments, sheet-like walls, and overdense nodes known as clusters, which bind vast underdense regions known as cosmic voids, typically with diameters of 10–15\,$h^{-1}$ Mpc \citep{2001ApJ...557..495P}. Although voids occupy a substantial fraction of the cosmic volume (approximately 70\%), they contain only around 10\% of the total matter content \citep{2014MNRAS.441.2923C}. 

Cosmic voids originate from primordial density fluctuations and evolve under weaker gravitational forces, allowing them to expand faster than the general Hubble flow \citep{2010MNRAS.404L..89A, 2011IJMPS...1...41V}. As a result, they are significantly less populated by galaxies and largely unaffected by environmental processes such as ram-pressure stripping and tidal interactions, which are prevalent in denser regions of the cosmic web \citep[e.g.][]{2013MNRAS.430.3017B, 2022A&ARv..30....3B}. Studying galaxy evolution under the unique environmental conditions of voids provides an opportunity to disentangle the role of the large-scale environment in shaping mass assembly (both baryonic and dark matter), regulating galaxy physical properties, and driving the transformation from star-forming to quiescent systems. Investigating these processes in cosmic voids constitutes the primary objective of Calar Alto Void Integral-field Treasury Survey \citep[CAVITY;][]{2024A&A...689A.213P}, an ongoing legacy programme designed to study void galaxies (VGs) across a range of nearby voids with different sizes and dynamical stages.

Prior to the CAVITY, studies of VGs were relatively limited in number and were largely based on small samples. Nevertheless, they provided the first evidence that galaxies in these environments may follow evolutionary pathways distinct from those in denser regions of the cosmic web.  For instance, several studies showed that VGs are generally small, late-type systems, with bluer colours, more recent star formation with some evidence for ongoing cold-gas accretion in several cases \citep[see also][]{2004ApJ...617...50R,2008ApJ...673..715C,2009ApJ...696L...6S,2011AJ....141....4K,2012AJ....144...16K,2016MNRAS.458..394B, 2017A&A...600L...6K, 2017MNRAS.464..666B}. More recently, others report that core-collapse supernovae (SNe) occur more frequently in VGs, whereas Type Ia SNe are comparatively rarer \citep{2026ApJ...997..121W}. These results suggest ongoing star formation and a relative deficit of older stellar populations in VGs. Numerical simulations likewise suggest systematic differences between VGs and galaxies in higher-density environments. For example, \cite{2020MNRAS.493..899H} reported an enhanced fraction of low-mass black holes in the inner regions of voids, while \cite{2024MNRAS.528.2822R} found, based on the {\,IllustrisTNG300} simulation, that VGs tend to experience mergers at later times, leading to a more recent build-up of their present-day stellar mass ($M_{\star}$).

However, the extent to which VGs differ from their counterparts in denser environments remains a matter of debate. For instance, analyses of low-mass galaxies in the Lynx-Cancer void reported systematically lower gas-phase metallicities than their counterparts in denser environments \citep{2011AstBu..66..255P}, whereas \cite{2015ApJ...798L..15K} did not find such a difference. Similarly, using SDSS fibre spectroscopy, \cite{2018ApJ...861..101M} found no significant differences in the central ($\sim$ 1 kpc) stellar velocity dispersions or stellar mass surface densities of quiescent VGs compared to their counterparts in denser environments. Alongside these discrepancies, several key aspects of the formation and evolution of VGs, including their mass assembly histories, dark matter content, and chemical enrichment, remain poorly constrained. Addressing these questions requires large, homogeneous samples and observations capable of mapping galaxies on spatially resolved scales. The CAVITY project was conceived to address this need and has since established the largest homogeneous integral-field spectroscopic (IFS) dataset of nearby VGs to date.

Since its inception, studies based on the growing CAVITY dataset have yielded important insights into several key aspects of void galaxy evolution. An analysis of VGs in the CAVITY sample of 15 voids, compared to galaxies in clusters and those not in clusters nor voids \citep[i.e. NCNV;][]{2026A&A...710A.337T, 2026arXiv260617676A}, uncovered a general bimodality in galaxy star formation histories (SFHs) across the cosmic web, separating systems into short-timescale (ST-SFH) and long-timescale (LT-SFH) evolutionary channels, while highlighting the slower mass assembly of VGs irrespective of their SFH class \citep{2023Natur.619..269D}. This delayed evolution is further reflected in the systematically lower stellar metallicities of VGs, and as \citet{2023A&A...680A.111D} showed, these differences are particularly pronounced among low-mass galaxies, systems with LT-SFHs, and spiral galaxies in voids. Similarly, \cite{2026arXiv260525557B} investigated the gas-phase metallicity of low-mass galaxies across the cosmic web and reported a similar trend, finding that VGs exhibit lower gas metallicities than their counterparts in clusters.

The picture of VGs delayed evolution was further reinforced by \citet{2024A&A...687A..98C}, who performed both global and spatially resolved analyses on stellar population properties of CAVITY galaxies including the IFS data released as the survey's first data release \citep[hereafter \citetalias{2024A&A...691A.161G}:][]{2024A&A...691A.161G}. Their study showed that VGs are, on average, younger and exhibit more delayed evolutionary pathways, particularly in the outer regions of their discs, where the star formation rate (SFR) is also slightly enhanced \citep[see also][]{2026A&A...709A.227C}. They further found that VGs possess lower stellar mass surface densities than their non-void counterparts, with these differences being most pronounced among low-mass systems \citep[see also][]{2024RMxAA..60..323S}. Together, these results consistently support a scenario of slower and more gradual evolution for galaxies residing in voids \citep[see also][]{2026ApJ...997..121W}.

Analyses of the mass-size relation in VGs provide further evidence for their less evolved nature. In particular, \citet{2025A&A...695A..84P} showed that early-type galaxies in voids are typically 10 to 20\% smaller than their counterparts in denser environments \citep[see also][]{2018ApJ...861..101M}. While no significant size differences are observed among massive late-type galaxies, this trend is reversed at lower stellar masses, where late-type VGs appear larger than their counterparts in clusters. These results further support a scenario of slower structural evolution in voids and suggest reduced minor-merger activity among early-type VGs. Moreover, such findings highlight the key role of large-scale environment in regulating baryonic mass assembly and morphological transformation. In low-density regimes, the reduced influence of external processes (e.g. mergers, ram-pressure stripping, and harassment) allows internal mechanisms, such as secular processes and feedback, to become primary drivers of galaxy evolution. This has already been discussed by \cite{2025A&A...693L..16B} for a rare sample of isolated and quenched dwarf VGs, identified using the CAVITY sample, whose evolution and quenching may have been driven primarily by internal feedback mechanisms.

Despite the common perception, not all the VGs are completely isolated. The work of \cite{2026arXiv260617676A} shows that galaxy groups are also present in void environments. These groups appear to be in the earlier stages of their dynamical evolution, and their richness does not depend strongly on void density or on the distance to the centre of the voids. However, these groups are generally smaller, since the most massive galaxy group identified in the CAVITY sample hosts only six members. \citet{2024A&A...691A.341T} highlighted the important role of groups and the local environment in voids, showing that VGs residing in groups tend to assemble their stellar mass earlier than isolated systems. Along with the local environment, the influence of the host dark matter halo and the surrounding large-scale structure is also required to explain the observed distributions of stellar mass and morphology \citep{2026A&A...710A.337T}.

The presence of galaxy groups and mergers within voids \citep[e.g.][]{2026A&A...706A.194A}, even in their central regions and in apparently isolated systems \citep[][]{2025A&A...698A.260B}, suggests that void dynamics is also a key parameter for understanding the evolution of both large-scale structure and the galaxies residing within them. This idea is further reinforced by \citet{2023A&A...673A..38C}, who investigated the total (dark and luminous) matter content of seven CAVITY voids and showed that the majority of these systems are not expanding in all three spatial dimensions. Instead, they may experience contraction along one or two dimensions, potentially driven by the tidal influence of the surrounding matter distribution.

Together, these studies illustrate the unique capabilities of the CAVITY project and motivate the continued expansion of its publicly available dataset. In this article, we present the second data release (DR2) of the CAVITY project, which doubles the initial sample from \citetalias{2024A&A...691A.161G} by providing public access to an additional 100 galaxies (200 in total). Following the DR1 steps, we describe the sample selection, observational strategy, data reduction pipeline, and quality-control procedures. Additionally, we provide a comprehensive characterisation of both the large-scale and local environments of the surveyed VGs, including the properties of the host voids and the group membership of individual galaxies. This information provides an important framework for interpreting the spatially resolved properties of VGs in the context of both their host voids and their immediate surroundings. This data release focuses exclusively on the IFS data products and does not include extensions of the CAVITY project \citep[i.e. CAVITY+;][]{2022A&A...658A.124D,2024A&A...689A.213P, 2024A&A...692A.125R, 2026arXiv260523399E}.

This paper is organised as follows: in Section~\ref{Cavity:parent_sample}, we briefly outline the criteria used to construct the CAVITY parent and observable samples, and in Section~\ref{Obs.Red.} we explain details of observation strategies and the data reduction phase. In Section~\ref{QC}, we describe the quality control procedure. In Section~\ref{Sample}, we present the CAVITY DR2 sample. This is followed by Section~\ref{Analysis}, where we characterise the voids and local environment of VGs in the CAVITY parent sample, compared to the DR2. In Section~\ref{Summary} we summarise the main conclusions. In this paper, we assume a flat $\Lambda$CDM
cosmology with H$_{0}$\,=\,70\, km~s$^{-1}$ Mpc$^{-1}$, $\Omega_{\rm M}$\,=\,0.30, and $\Omega_{\rm \Lambda}$\,=\,0.70. 

\section{CAVITY sample selection}\label{Cavity:parent_sample}
The CAVITY parent sample, from which the survey targets are selected for observation, is constructed after applying a set of selection criteria to the SDSS void catalogue of \citet{2012MNRAS.421..926P}, which originally identified 1055 voids in $z<0.107$. In particular, CAVITY focuses on the redshift range $0.005~\leq~z~\leq~0.050$, ensures full coverage of voids within the SDSS footprint, and concentrates on voids containing at least 20 galaxies. {The latter criterion is applied to ensure sufficient galaxy statistics for a robust characterisation of the void size and galaxy distribution. }These criteria reduce the initial sample to 42 voids, from which a representative subset of 15 voids is selected to span a broad range of void and galaxy properties (e.g. redshift, size, and $g$-$r$ colour distribution), while also satisfying visibility constraints from the Calar Alto Observatory \citep[for more details see][]{2024A&A...689A.213P}. The CAVITY parent sample comprises 4866 galaxies distributed across these 15 voids.

To minimise contamination from other components of the large-scale structure, objects overlapping with the clusters \citep[defined as groups of at least 30 galaxies;][]{1989ApJS...70....1A} in \cite{2017A&A...602A.100T} catalogue are excluded, and only galaxies located within 0.8 effective radius of a void ($R_\mathrm{e, void}$) are retained. Here, $R_\mathrm{e, void}$ is defined as the radius of a sphere with the same volume as the actual void \citep[see][]{2012MNRAS.421..926P, 2024A&A...689A.213P}. This selection ensures minimal influence from surrounding filamentary structures and prioritises galaxies in the innermost regions of voids. Furthermore, galaxies with face-on or edge-on orientations introduce additional complexities in spectral analyses, particularly for gas and stellar kinematics. Therefore, to ensure a proper analysis, only galaxies with intermediate inclinations (20 to 70~degrees) are considered \citep{2024A&A...689A.213P}. These criteria yield a sample of 1115 observable VGs (hereafter the observable sample), from which the data release samples, including the 200 VGs of DR2 are drawn. The distribution of the CAVITY parent sample and the observable sample in the colour–magnitude diagram is shown in Fig.~\ref{fig:sample_CMD} together with that of DR2 galaxies.

\section{Observations and data reduction}\label{Obs.Red.}
Between its commencement in January 2021 and May 2026, CAVITY was assigned a total of 297 observing nights with the 3.5-m telescope at the Calar Alto Observatory using the Potsdam Multi-Aperture Spectrophotometer \citep[PMAS;][]{2005PASP..117..620R} in PPAK mode. By the end of this period, observations had been completed for 307 unique galaxies, some of which were observed in multiple observing runs.

\subsection{Observations }\label{Obs.Red.:observations}
PMAS/PPAK consists of a single hexagonal bundle of 331 fibres, each with a diameter of 2.7\arcsec, providing a field of view (FOV) of $74~\times~64$~arcsec$^{2}$. An additional set of six mini-bundles, comprising a total of 36 fibres, is arranged in a ring at a radius of $\sim80$~arcsec from the centre of the FOV to sample the sky background. All CAVITY DR2 galaxies were observed with the setup of V500 grating, which provides a spectral resolving power of $R \sim850$ at $\sim5000$\,\AA\ and a full width at half-maximum (FWHM) of $\sim6$\,\AA. This setup covers the key optical spectral features, spanning the wavelength range 3745-7500 \AA.

Exposure times were determined following the same strategy as in \citetalias{2024A&A...691A.161G}. In brief, galaxies were divided into two categories, bright and faint, based on elliptical isophotal profiles fitted to their SDSS DR12 $r$-band photometric data. Galaxies with a surface brightness $\mu > 22.1$ mag arcsec$^{-2}$ at the effective radius were classified as faint and observed with a total integration time of 3 hours, while those with $\mu \leq 22.1$ mag arcsec$^{-2}$ were classified as bright and observed with a total integration time of 1.5 hours. {We adopted the SDSS $r$-band \texttt{PetroR50}, defined as the radius enclosing 50\% of the $r$-band Petrosian flux} \citep{2015ApJS..219...12A}, as a proxy for the effective radius ($R_{\mathrm{e}}$). All the VGs were observed using the same three-position dithering scheme as described in \citetalias{2024A&A...691A.161G}, designed to improve spatial sampling, mitigate vignetting effects (see Section~\ref{QC}), and enhance the spatial coverage of the central bundle. At each dithering position, two and four exposures of 900\,s were obtained for bright and faint galaxies, respectively. In addition, one set of calibration frames for bright galaxies, and at least two sets for longer exposures (faint systems) were obtained.

\begin{figure}
\centering
    \includegraphics[width=0.95\columnwidth]{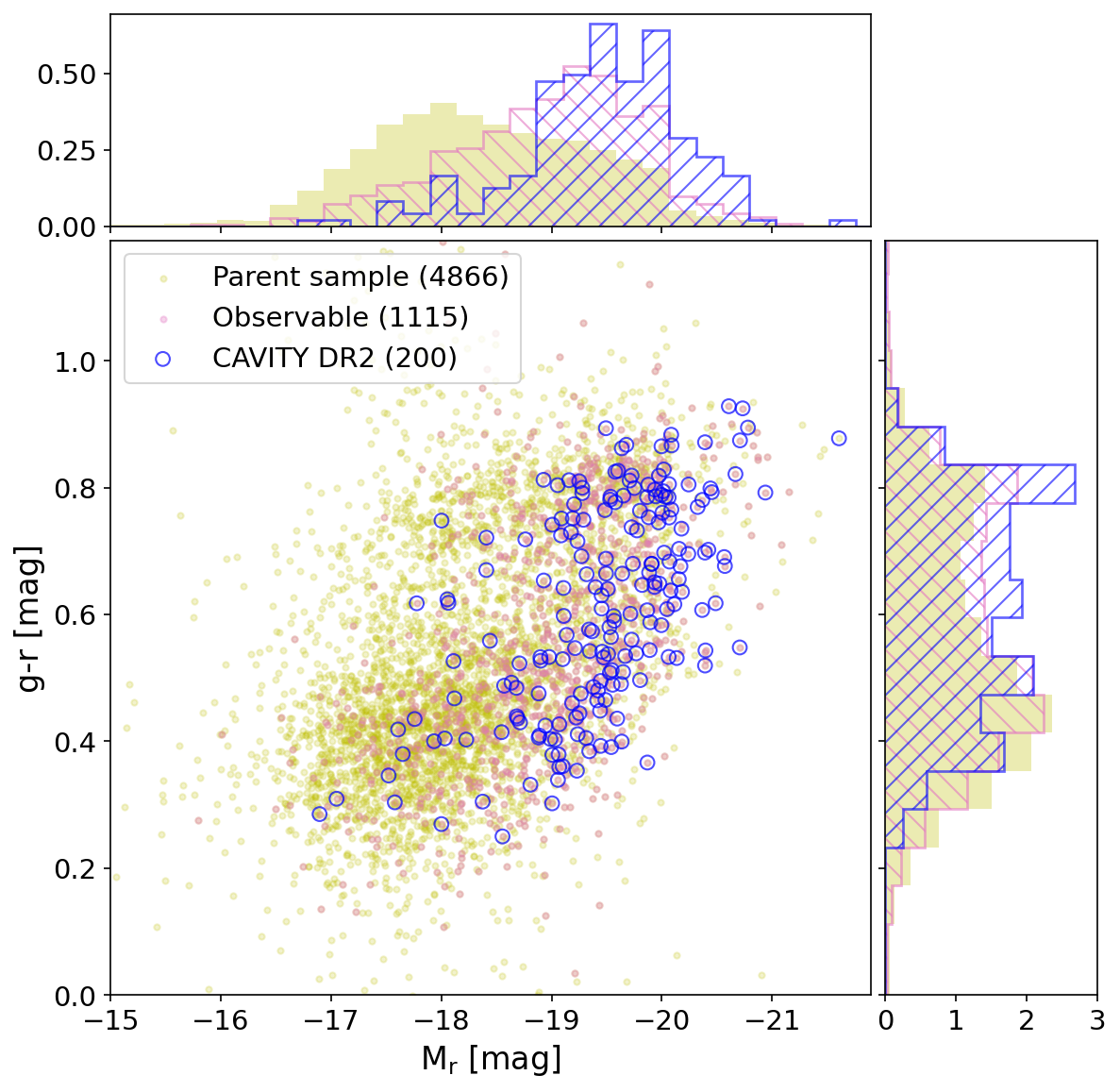}
\caption{Distribution of CAVITY galaxies in the SDSS $g-r$ vs. $M_r $ colour-magnitude diagram. The colour and absolute magnitude distributions are displayed in the top and side panels, respectively. Yellow dots (histograms) denote galaxies in the CAVITY parent sample (4866 galaxies), while pink and blue are used for the observable subsample (1115) and the final CAVITY DR2 sample (200), respectively.}  \label{fig:sample_CMD}
\end{figure}

Given the relatively long exposure times required for each galaxy, together with the shared operation of multiple observing programmes at the observatory, a fraction of the VGs, including part of the DR2 sample, was necessarily observed over multiple nights. To maintain data cube quality and minimise uncertainties related to fibre positioning and atmospheric transparency, we adopted the same strategy as in \citetalias{2024A&A...691A.161G}, requiring each individual dithering position to be completed within a single night. Of the DR2 sample, 66\% of the galaxies were completed within a single night, 30\% over two nights, and the remainder over three
nights. Similarly, pointings that do not satisfy the minimum quality-control requirements (see Section~\ref{QC} and the \citetalias{2024A&A...691A.161G}) were rescheduled and re-observed to ensure the delivery of high-quality data cubes. 

\subsection{Data reduction}\label{Obs.Red.:Reduction}

The data reduction for CAVITY DR2 was carried out using the same automated pipeline employed for \citetalias{2024A&A...691A.161G}, implemented in version 1.2 of the CAVITY reduction framework. The pipeline operates with minimal human intervention and produces science-ready data products together with quality-control metrics that allow assessment of the reliability of the processed data. Its design follows the reduction procedures originally developed for the CALIFA survey \citep{2012A&A...538A...8S,2013A&A...549A..87H}, with adaptations tailored to the instrumental configuration and observing strategy of CAVITY. The pipeline is implemented in Python 3, with selected computationally intensive tasks optimised using Cython \citep{2011CSE....13b..31B}. For DR2, the same pipeline version 1.2 was used, with only minor internal adjustments aimed at improving computational efficiency and automation in specific cases. These modifications facilitate a smoother workflow but do not introduce any changes in the scientific treatment of the data.

The reduction begins with the processing of the raw CCD frames. Depending on the observing period, exposures were obtained using either the original four-amplifier readout configuration or the two-amplifier mode adopted after the detector intervention in 2022  (see \citetalias{2024A&A...691A.161G} for details on this issue). In both cases, the individual files produced by the detector readout are merged into a single frame per exposure. During this stage, the pipeline propagates the relevant sources of uncertainty, including Poisson and readout noise, while accounting for defective pixels and columns. Pixel cleansing is applied to remove cosmic rays, and additional corrections are performed to account for instrumental effects such as vignetting and stray light.

A stray light map is reconstructed from the inter-fibre gaps on the detector and subtracted from both calibration and science exposures. The spectra are then extracted from the detector images using an optimal extraction algorithm that incorporates the measured fibre profiles and allows proper propagation of pixel-level uncertainties. The extracted spectra are stored in row-stacked-spectrum (RSS) format, which constitutes the basic intermediate data product used in subsequent processing steps.

Wavelength calibration is derived from arc-lamp exposures acquired as part of the calibration sequences, allowing the determination of the wavelength solution and spectral resolution for each fibre. The spectra are subsequently resampled onto a common linear wavelength grid, incorporating corrections for small flexure offsets between calibration and science frames. Bad pixels are identified and masked during this process, and the corresponding uncertainties are propagated accordingly. Relative fibre-to-fibre transmission variations are corrected using twilight sky exposures that provide a uniform illumination reference across the instrument.

Flux calibration is performed using a master sensitivity curve derived from repeated observations of spectrophotometric standard stars. Science frames are first corrected for atmospheric extinction using the monitored extinction values at the observatory and the airmass of the observations. The sensitivity curve is then applied to convert detector counts into flux units. After flux calibration, the sky contribution is removed using the dedicated 36 sky fibres of the PPAK bundle, which sample the night sky simultaneously with the science observations.

Finally, the three dithered exposures obtained for each target are combined and spatially resampled to reconstruct the final data cube. This reconstruction uses a flux-conserving interpolation scheme based on inverse-distance weighting and includes an empirical correction for differential atmospheric refraction \citep{Shepard:1968}. The resulting cubes are subsequently adjusted to match the absolute photometric scale using broadband SDSS imaging. As in \citetalias{2024A&A...691A.161G}, the spatial interpolation introduces correlated noise between adjacent spaxels. This effect is characterised empirically and parameterised following the relation provided in \citetalias{2024A&A...691A.161G} (Eq. 1), ensuring proper error estimation when combining spectra. Overall, the reduction applied to DR2 is identical to that used for \citetalias{2024A&A...691A.161G}, ensuring full consistency between both releases while benefiting from minor improvements in efficiency and automation within the pipeline workflow. The final reduced data for each galaxy are delivered as a single file, following the same standard binary FITS format and structure adopted in \citetalias[][]{2024A&A...691A.161G}.

\section{Data quality control}\label{QC}
The full quality-control (QC) procedure was applied to all galaxies observed and reduced up to March 2026, ensuring that the released products met the standard requirements for scientific use by the community. As in DR1, the CAVITY QC procedure consists of two complementary stages: a visual inspection and an automated assessment.

The visual inspection of each final reduced data cube was performed by the QC team with 23 members using a set of diagnostic plots automatically generated following data reduction. The inspection was carried out with \texttt{GALAssify}\footnote{\url{https://astrogal.gitlab.io/GALAssify/}}, a Python package developed for the visual classification of astronomical data products \citep{alcazar2025galassify}. The inspection was based on different quality flags that probe key aspects of cube integrity, including reconstruction artefacts, spatial distortions, spectral anomalies, and vignetting effects. Each galaxy was independently inspected by two members of the QC team. In cases of disagreement, a third reviewer's assessment was used to resolve the discrepancy. Flags identified by at least two reviewers are considered confirmed issues for the corresponding cube. During the visual inspection, key aspects of cube integrity were assessed. In particular:
\begin{itemize}
    \item Cube reconstruction, including artefacts associated with the dithering scheme, contamination from bright stars or nearby galaxies, and targets with surface brightness too low to be adequately sampled even after spatial binning. These aspects are inspected using white-light images constructed by integrating the cubes over the wavelength range 4500-7000 \AA, maps of negative-value spaxels to identify low-transmission fibres, and Voronoi-binning maps \citep{Cappellari2003} with a target S/N$>$30, used to assess the spatial extent of the galaxy that can be reliably analysed after adaptive binning for a given exposure time and galaxy surface brightness.
    \item Spectral quality, including distortions, atypical continuum shapes, spectral artefacts, and vignetting effects. These are assessed using the central spectrum together with integrated spectra extracted within 0.5, 1, and 2 $R_{\mathrm{e}}$. Vignetting refers to the reduction in transmission affecting those portions of the fibres that are projected onto the four corners of the CCD, and is a known limitation of the telescope's optical system. In reconstructed cubes, this appears as an annular region located at approximately 15 arcsec from the FOV centre and typically manifests as discontinuities at the spectral edges. In severe cases, vignetting can reduce the usable wavelength range to 4240-7140 \AA, although its impact on the CAVITY sample is generally limited, as VGs are typically compact (with isophotal diameter\footnote{Following the convention adopted in CAVITY, the optical diameter was estimated as 3$\times R_{90,r}$, where $R_{90,r}$ is the SDSS r-band Petrosian radius.} $d_{25}<40$ arcsec) and only their outermost regions are affected. To minimise this effect, all observations were conducted under the requirement that the target be accurately centred within the FOV.
    \item A previously reported issue associated with CCD quadrant ‘a’, which produced a sharp gradient in sky level between the northern (top) and southern (bottom) parts of the FOV, as well as a discontinuity in the spectra at $\sim$5800 \AA\ \citepalias[for full details see][]{2024A&A...691A.161G}. Although this issue was resolved during the preparation of DR1 and all affected galaxies were re-observed, this check has been retained in the visual QC procedure to ensure that similar artefacts do not occur again. As in DR1, this assessment is based on two diagnostic plots: (i) a zoomed-in view of the integrated spectra within 0.5, 1, and 2 $ R_{\mathrm{e}}$, focusing on the affected spectral range (5500–6100 \AA); and (ii) a comparison of the mean flux in each row of spaxels across the field of view from north to south, after masking the central region dominated by galaxy light.

\end{itemize}

\begin{figure*}
\centering
    \includegraphics[width=2\columnwidth]{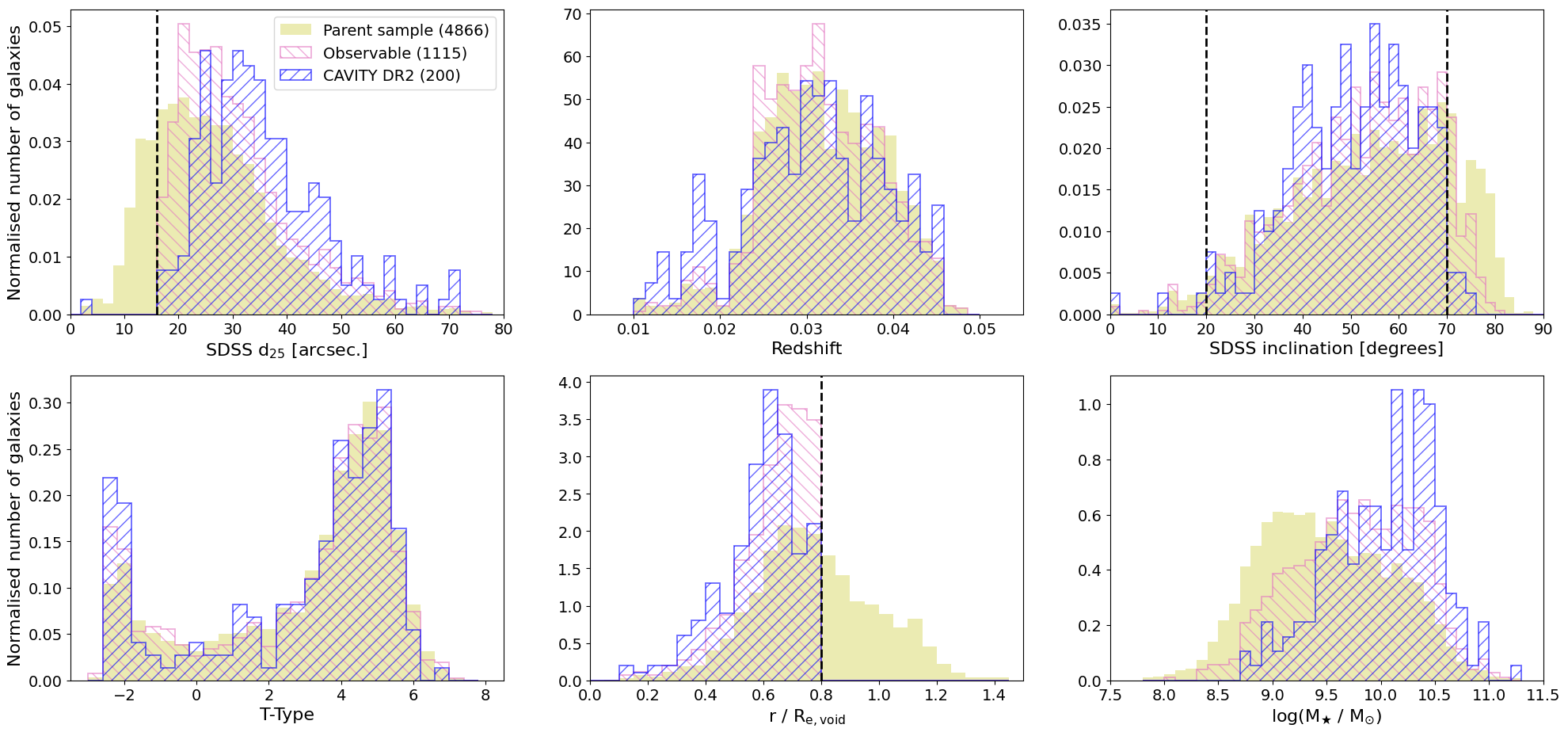}
\caption{Characterisation of the DR2 CAVITY sample in comparison with the CAVITY parent sample and observable sample from which observed galaxies are drawn (see Section~\ref{Sample} for details). From top to bottom and left to right, we show the distribution of $d_{25}$, the redshift, the inclination from SDSS, the morphological T-Type, the normalised distance to the centre of the void and the stellar mass. In solid light-green are the histograms corresponding to the CAVITY parent sample, in pink (left-slanting stripes) the observable sample, and in blue (right-slanting stripes) the final CAVITY DR2 sample. To the right of the black vertical dashed line in the normalised distance to the centre of the void panel histogram (bottom-left) we find the most external parts of the void (area between 0.8 and 1.0 times the void effective radius). This region is avoided as it can be considered as a transition zone from the void to the filamentary-like environment. Galaxies that appear to fall outside the adopted CAVITY selection limits (e.g., in $d_{25}$ or inclination) reflect the use of the original SDSS catalogue values in these plots. These parameters were subsequently revised following the visual quality-control process and therefore do not necessarily correspond to the final values adopted by CAVITY.}  \label{fig:sample}
\end{figure*}

From the list of newly observed galaxies, only those that passed the visual inspection phase form the rest of the DR2 candidate list. These candidates were subsequently evaluated through the automated quality-control procedure, based on a series of quantitative parameters designed to characterise different aspects of data-cube quality. For this purpose, in \citetalias{2024A&A...691A.161G} we defined a set of thresholds and associated flags for each parameter (see Table~4 of the \citetalias{2024A&A...691A.161G} paper), and the same criteria are adopted here. These flags are generally based on the mean, maximum, and standard deviation of the corresponding parameters, as outlined below. Most of the automated QC flags are treated as warnings and do not necessarily indicate that a data cube is unsuitable for scientific use; rather, they highlight minor issues that may slightly affect data quality. The parameters considered are grouped into three main categories:
\begin{enumerate}
    \item The quality of the observing conditions is assessed using three primary parameters: airmass, sky brightness, and atmospheric extinction. All CAVITY observations are acquired at an airmass below 1.4 (corresponding to an elevation over $45^\circ$), well within the warning threshold defined in DR1. Consequently, only the standard deviation of the airmass is considered for quality assessment, with values exceeding 0.15 raising a warning; this occurs for fewer than 10\% of the candidate galaxies. The sky brightness in the $V$ band is estimated using the 30 faintest sky fibres in each pointing, and both the mean and standard deviation of these measurements are evaluated against the adopted warning thresholds. None of the galaxies exceed the threshold for mean sky brightness (19.5 mag arcsec$^{-2}$), while 53 candidate galaxies surpass the threshold of 0.1 mag arcsec$^{-2}$ for the standard deviation of the sky background (see \citetalias{2024A&A...691A.161G} for details). Atmospheric transparency is characterised through the observatory’s continuous monitoring of the $V$-band extinction for each observed pointing. Based on the mean, maximum, and standard deviation of the measured extinction, warnings are raised for 29\% of candidate galaxies, indicating non-uniform observing conditions.
    \item The quality of the instrumental performance and data reduction was assessed using four main diagnostics: stray light contamination, spectral dispersion, cross-dispersion, and residuals from the subtraction of bright sky lines. To ensure that excessive stray light residuals do not compromise the data cubes, strict thresholds were imposed on the mean, maximum, and standard deviation of this parameter (see Table~4 of the \citetalias{2024A&A...691A.161G} paper). However, none of the DR2 candidates surpassed these thresholds, and therefore, no warnings were raised because of this. The spectral dispersion and cross-dispersion were measured using arc-lamp calibration frames and continuum-lamp exposures, quantified as the FWHM of the spectral lines and fibre traces, respectively. Five and four galaxies in the DR2 candidate sample raised warnings by exceeding the thresholds defined in DR1 for these two parameters, respectively. The accuracy of sky subtraction was assessed by measuring the minimum and maximum flux residuals of the 5577\AA\ [OI] sky line across all pointings, based on fibre-averaged residual spectra. In addition, the maximum standard deviation of these residuals was considered. Applying the same thresholds adopted in DR1, warnings were raised for 85\% of the candidate sample owing to elevated residual levels and/or large residual scatter. However, similar to DR1, these values do not exceed 0.5 and 1.6 in maximum and standard deviation, respectively. 
    \item The quality of the calibrated data products was assessed by inspecting the global spectrophotometric calibration and the stability of the wavelength calibration across the spectral range. As described in Section~\ref{Obs.Red.:Reduction}, CAVITY data cubes are rescaled to match the absolute flux level of the SDSS DR16 broad-band photometry \citep{2020ApJS..249....3A} measured within a 30-arcsec diameter aperture. This flux calibration can be affected by several factors, including the residual impact of vignetting in a small fraction of fibres, the partial coverage of the circular aperture in the DR16 image, as well as artefacts and minor contamination from foreground stars in the SDSS DR16 imaging. To quantify the accuracy of the flux calibration, the median scaling factors were required to lie within the range 0.7–1.6. Galaxies with values outside these limits were excluded from the final sample.
\end{enumerate}

Out of the 194 newly observed galaxies available by March 2026 that went through the QC process, those failing to pass the visual inspection stage of the QC were excluded from the present data release (71 galaxies). A detailed inspection of these systems revealed that the reported issues were primarily associated with observational artefacts (e.g. satellite trails) or problems introduced during the reduction and calibration processes. These issues will be revisited and, where possible, corrected in preparation for the next data release. Among the remaining galaxies, 12 systems that passed the initial visual inspection failed the automatic QC stage, mainly due to unreliable flux calibration caused by partial coverage of the galaxies in the SDSS imaging or the presence of fibres with reduced throughput. These cases will also be reassessed for the next data release. 

To provide a seamless extension of \citetalias{2024A&A...691A.161G}, the 100 galaxies released as part of the first data release were retained unchanged in DR2. The remaining 100 galaxies were randomly selected from the pool of 111 newly available high-quality data cubes, maintaining the planned total sample size of 200 galaxies for this release. The remaining 11 science-grade data cubes will be made publicly available as part of the next CAVITY data release.

\section{CAVITY DR2 Sample}\label{Sample}
As shown in Fig.~\ref{fig:sample_CMD}, overall, the CAVITY DR2 sample follows a distribution similar to that of both the observable and parent samples. Compared to the previous data release, DR2 mitigates the slight excess of galaxies at the bright end (brighter than the absolute r-band magnitudes, $M_{r}$, of $-$20~mag) by including a larger fraction of fainter systems. Nevertheless, the low-mass, low-luminosity regime (fainter than $M_r = -18$~mag) remains largely inaccessible with the current observational setup. The low-mass regime (in particular $\log(M_\star/\rm M_\odot)~<~9.5$) is therefore underrepresented in the current sample and will be complemented by ongoing and future observing programmes within the CAVITY project.

\begin{figure}
\centering
    \includegraphics[width=\columnwidth]{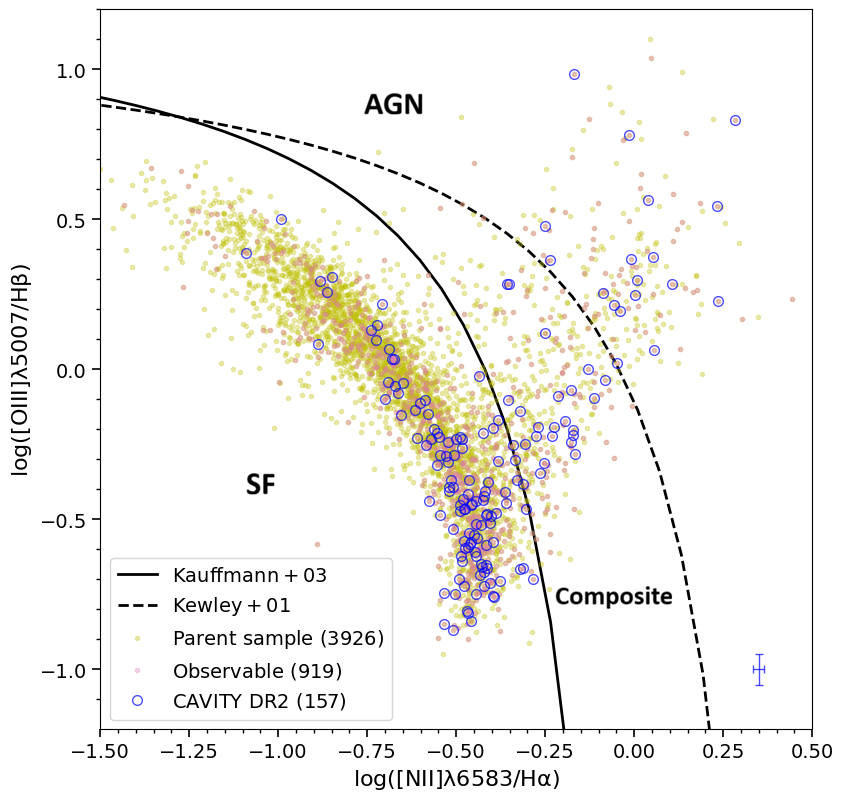}
\caption{Diagnostic diagram (BPT) of [OIII]$\lambda$5007/H$\beta$ versus [NII]$\lambda$6584/H$\alpha$. The typical error in each axis for DR2 is represented with a blue cross in the bottom right of the figure. The dashed, continuous, and dotted lines show the \citet{2001ApJ...556..121K} and the \citet{2003MNRAS.346.1055K} demarcations, respectively.  Yellow dots, pink dots, and blue circles represent galaxies in the CAVITY parent sample (3926 galaxies), observable subsample (919), and the final CAVITY DR2 sample (157), respectively (with sufficient S/N in all diagnostic lines).}  \label{fig:BPT}
\end{figure}

Fig.~\ref{fig:sample} provides an overview of the main physical properties of the CAVITY DR2 sample in comparison with the observable and parent samples. The top left-hand panel shows the distribution of $d_{25}$, defined for each galaxy as three times its SDSS $r$-band \texttt{PetroR90}. Only galaxies with $d_{25}>16$~arcsec are considered observable, ensuring sufficient spatial sampling for two-dimensional analyses of their properties. All selected systems are fully contained within the PMAS/PPAK field of view ($\sim$70~arcsec). The top-middle panel shows the redshift distribution. The CAVITY parent sample, and consequently the observable and DR2 samples, are defined within the redshift range $0.005 \leq z \leq 0.050$, chosen to ensure adequate volume coverage of large voids. As shown in the top left-hand panel of Fig.~\ref{fig:sample}, only galaxies with intermediate inclinations (20–70 degrees) are considered observable and included in the DR2 sample, with the majority of DR2 galaxies lying between 35 to 70 degrees.

The distribution of the morphological type (T-Type) of the DR2 sample is consistent with that of the parent and observable samples, although it exhibits a modest excess of early-type galaxies. This bias arises naturally from the survey selection criteria, as massive galaxies are more likely to satisfy the observational requirements \citep[bottom left-hand panel; classifications from][]{2018MNRAS.476.3661D}. As discussed in Section~\ref{Cavity:parent_sample}, CAVITY prioritises the observation of galaxies located in the inner regions of voids for two main reasons: (i) to minimise contamination from surrounding large-scale structure, such as NCNV, and (ii) to probe galaxy evolution in the lowest-density environments of the cosmic web. To this end, galaxies in the observable sample are ranked according to their void-centric distance (bottom middle panel), and those with the smallest distances are selected first for observation within each void. 
The number of galaxies increases with void-centric distance up to to the imposed limit of 0.8 $R_{e,\text{void}}$. Lastly, the bottom right-hand panel of Fig.~\ref{fig:sample} compares the stellar mass distributions of the parent, observable, and DR2 samples.

Figure~\ref{fig:BPT} shows the distribution of galaxies from the CAVITY parent sample (3926 galaxies; yellow points), the observable subsample (919 galaxies; pink points), and the final CAVITY DR2 sample (157 galaxies; blue circles) on the BPT diagnostic diagram \citep{1981PASP...93....5B}. The sample sizes shown here are smaller than those reported in Fig.~\ref{fig:sample_CMD}, as only galaxies with positive emission-line fluxes and signal-to-noise ratios greater than or equal to 3 in all four diagnostic lines [OIII]$\lambda$5007, H$\beta$, [NII]$\lambda$6584, and H$\alpha$ are included. The emission-line measurements and stellar masses are taken from the Max Planck Institute for Astrophysics and Johns Hopkins University (MPA--JHU) catalogue, derived from fits to the SDSS fibre spectra \citep{2003MNRAS.341...33K,2004MNRAS.351.1151B,2004ApJ...613..898T,2007ApJS..173..267S}.

\begin{table*}
    \centering
    \begin{tabular}{c | c c c c c c c}
    \hline\hline  \\[-2ex]
        Sample & Total & MPA-JHU & S/N & Star-forming & Composite & AGN & Non-emitting galaxies\\
        (1) & (2) & (3) & (4) & (5) & (6) & (7) & (8) \\[0.5ex]
        \hline\\[-2ex]
        Parent & 4866 & 4562 & 3926 & 3361 (73.7\%) & 367 (8.0\%) & 198 (4.3\%) & 636 (13.9\%) \\
        Observable & 1115 & 1059 & 919 & 728 (68.7\%) & 118 (11.1\%) & 73 (6.9\%) & 140 (13.2\%) \\
        DR2 & 200 & 190 & 157 & 115 (60.5\%) & 24 (12.6\%) & 18 (9.5\%) & 33 (17.4\%) \\
        \hline\\[-2ex]
    \end{tabular}
    \caption{Summary of the total number of galaxies in each of the samples. The columns correspond to: 
(1) Sample of galaxies; 
(2) Total number of galaxies in the corresponding sample;
(3) Galaxies in the MPA-JHU catalogue;
(4) Galaxies satisfying the condition S/N $\geq$ 3 and positive fluxes;
(5) Star-forming galaxies;
(6) Composite galaxies;
(7) AGN galaxies;
(8) Passive or non-emitting galaxies. 
Percentages are calculated relative to column 3. 
}
    \label{tab:tab1}
\end{table*}

The parent and observable samples are dominated by star-forming galaxies, with the majority of objects (73.7\% and 68.7\%, respectively) lying below the empirical demarcation line of \citet{2003MNRAS.346.1055K}. The DR2 sample follows the same trend, with 60.5\% of galaxies classified as star-forming systems. A further 9.5\% of galaxies lie above the theoretical demarcation of \citet{2001ApJ...556..121K}, indicating that active galactic nuclei (AGN) dominate their ionisation budget. The remaining $\sim$ 12.5\% of the DR2 sample occupy the region between the two demarcations, commonly referred to as the composite region, where both star formation and AGN activity may contribute to the observed emission-line ratios.
In Table~\ref{tab:tab1} we show a summary of the number of galaxies in each of the steps described above (from columns 1 to 4). Furthermore, columns 5, 6, 7, and 8 indicate the number of galaxies (and their percentage relative to column 3) that are star-forming, composite, AGN, and non-emitting, respectively.

\begin{figure*}
\centering
    \includegraphics[width=1\textwidth]{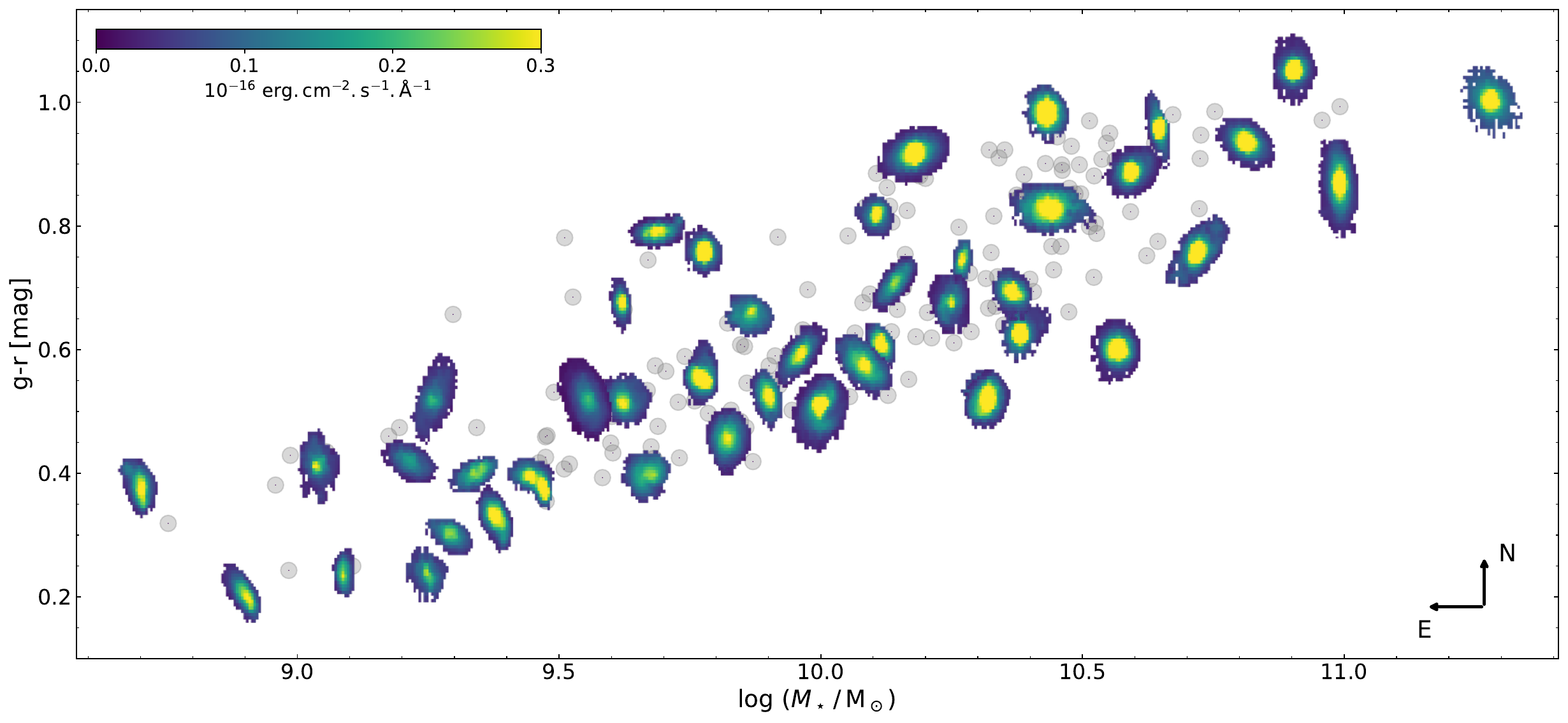} \\  
\caption{Moment 0 maps reconstructed from the CAVITY data cubes for 47 representative galaxies from the DR2 sample, overlaid on the colour-mass diagram. The Moment 0 maps were constructed by summing the integrated light over the full wavelength range observed. Spaxels with signal-to-noise ratio less than 3 are discarded. The distribution of the complete DR2 sample is shown by the grey data points.}  \label{fig:sample_moment0}
\end{figure*}

Fig.~\ref{fig:sample_moment0} shows the colour-mass diagram for the DR2 galaxies. For a representative sample of 47 galaxies, {we present the corresponding Moment 0 maps, which show the wavelength-integrated flux at each spaxel across the data cube's spectral range, providing a two-dimensional view of the galaxy's spatial light distribution. The maps are constructed using the Spectral-cube package\footnote{https://github.com/radio-astro-tools/spectral-cube}. To define the spatial extent of the galaxy in the moment 0 maps, we selected spaxels with mean S/N $>$ 5, measured over 100 \AA\, windows between 3900 and 6700 \AA. We then applied the convex hull of the resulting S/N-selected region to obtain a continuous spatial mask and exclude low-S/N regions outside the galaxy}. These examples provide a visual overview of the reconstructed data products and illustrate the diversity of galaxy morphologies across the stellar mass range covered by DR2. Instructions for accessing the CAVITY DR2 data products through the survey database are provided in Appendix~\ref{appendix:Data_access}.

\begin{figure*}
\centering
    \includegraphics[width=0.8\textwidth]{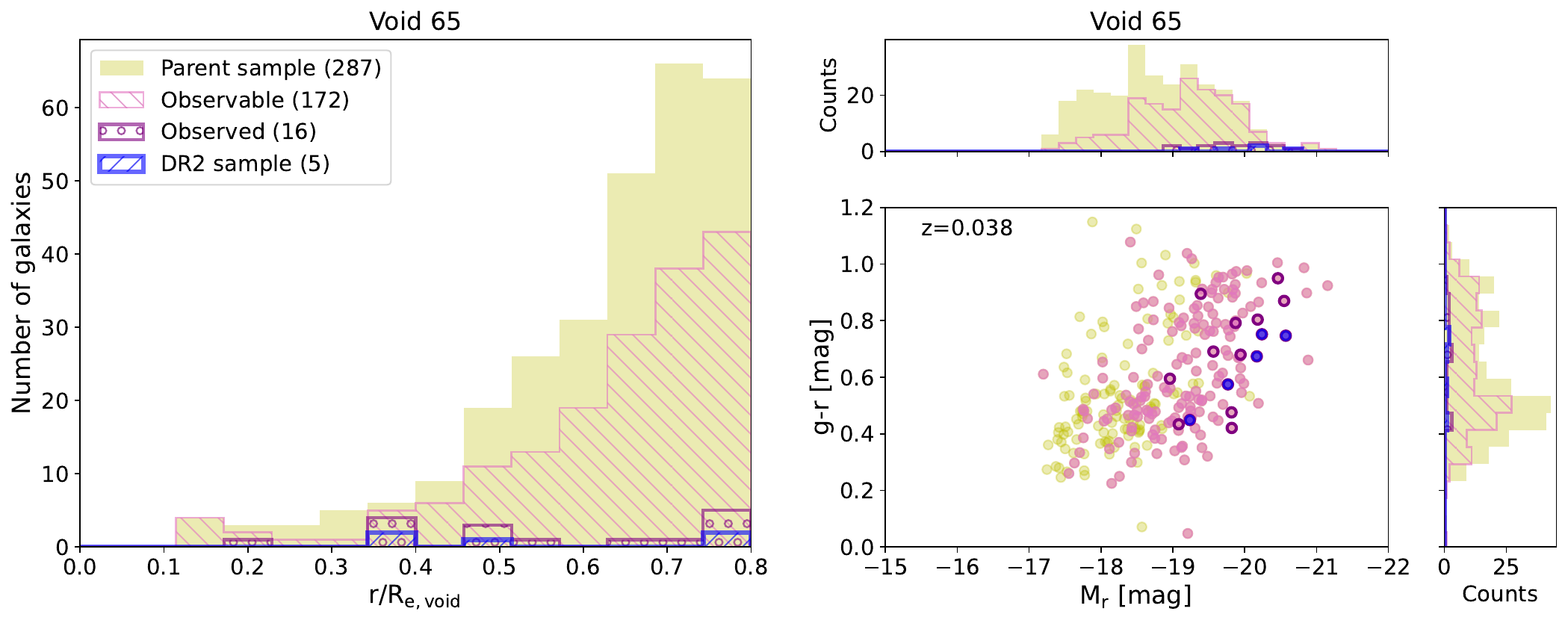}
    \includegraphics[width=0.8\textwidth]{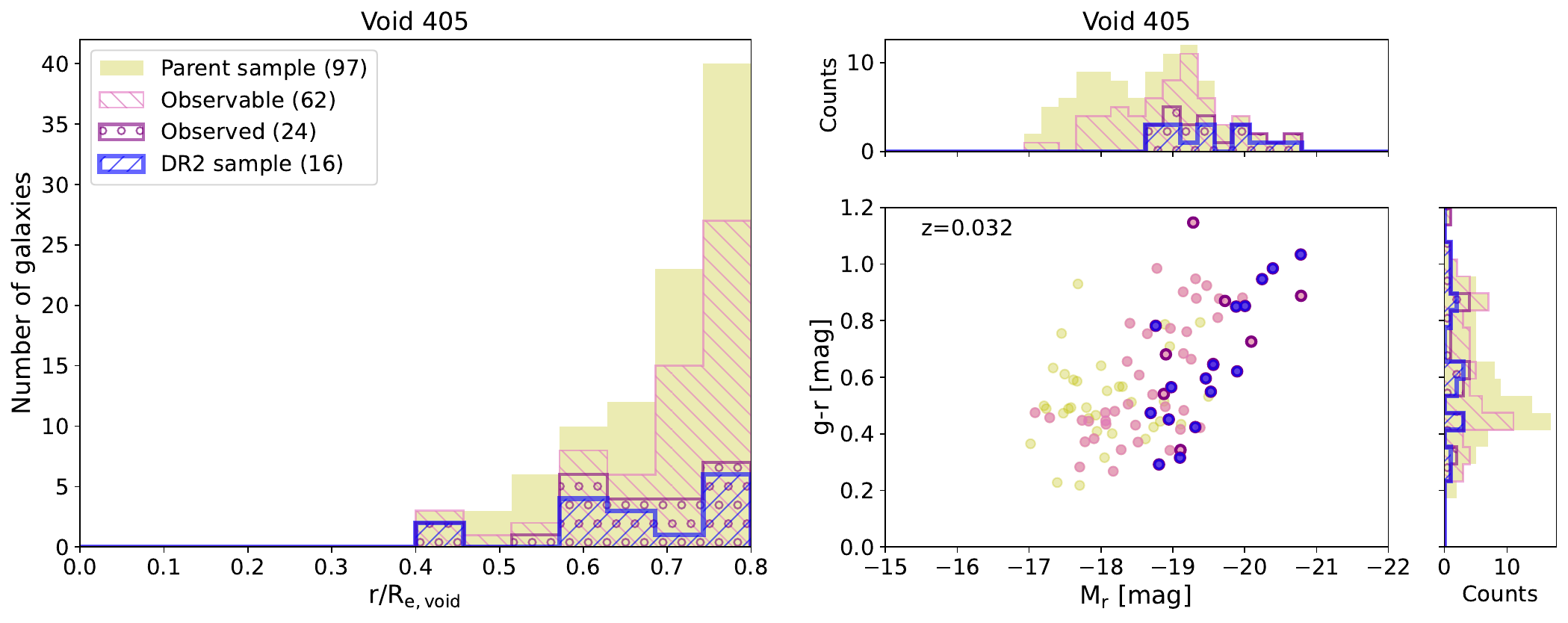}
    \includegraphics[width=0.8\textwidth]{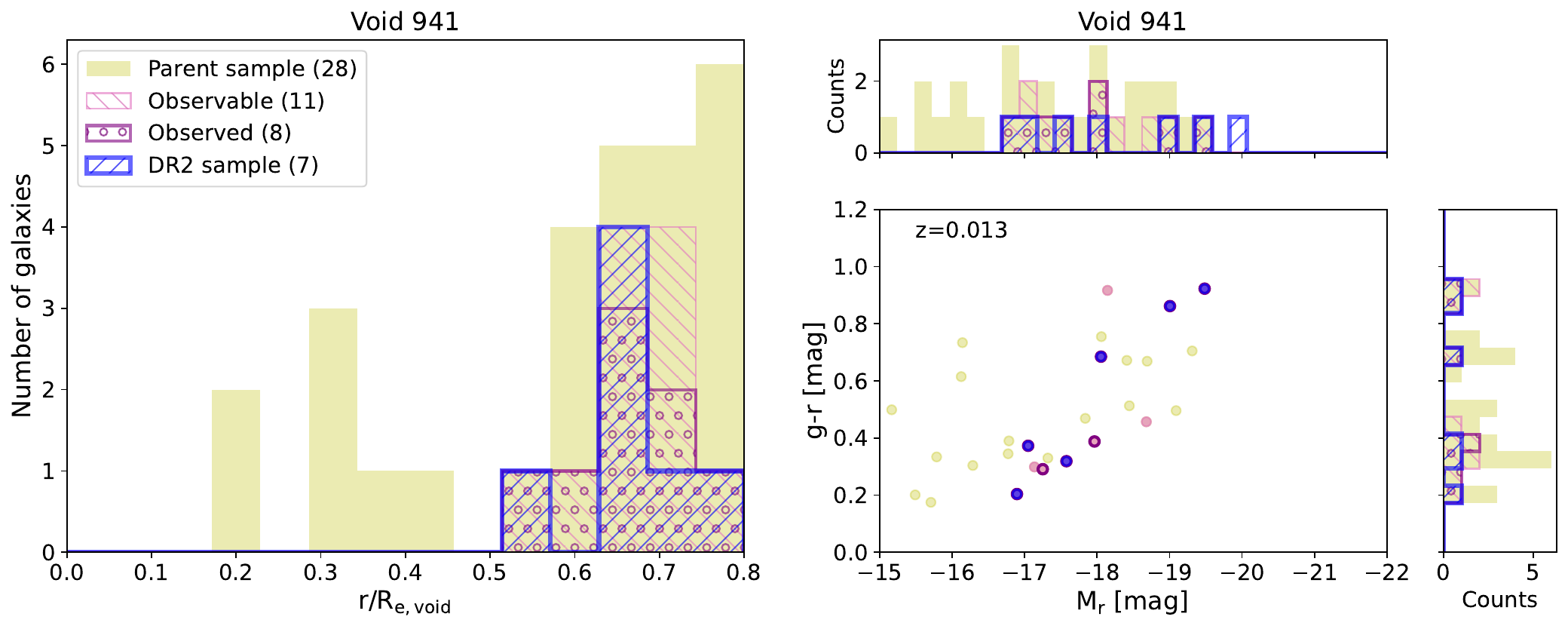}

\caption{The distribution of VGs in three example CAVITY voids. \textit{Left-hand panels:} The distribution of void-centric distances, as a function of $R_{\mathrm{e,void}}$ for all galaxies in, from top to bottom, voids 65, 405, and 941, respectively. \textit{Right-hand panels:} The colour-magnitude diagram for VGs in three voids. The top and right-hand histograms show the distribution of $M_r $ and $g$-$r$, respectively. In all panels, galaxies belonging to the CAVITY parent sample are shown in yellow, those in the observable sample in pink, galaxies with IFS data observed up to May 2026 in purple, and the CAVITY DR2 sample in blue. The redshift quoted in each panel is the median redshift of all galaxies belonging to the corresponding void.}  \label{fig:Voids}
\end{figure*}

\section{The environmental characterisation of the DR2 sample}\label{Analysis}
Despite defining the vast, low-density regions of the cosmic web, voids are far from uniform \citep[e.g.][]{2023A&A...673A..38C}. Their galaxy populations likewise exhibit significant diversity in both distribution and physical properties. Furthermore, although VGs reside in low-density environments, this does not necessarily imply that they are all isolated or free from interactions with nearby systems. Indeed, \cite{2026arXiv260617676A} demonstrated that galaxy groups, triplets, and pairs are also present in cosmic voids, although they are generally smaller and appear to be at earlier stages of dynamical evolution than their counterparts in denser environments \citep[see also][]{2026A&A...710A.337T}. 

To better characterise the environment of CAVITY DR2 galaxies, in Section~\ref{Analysis:voids} we first review the main properties of the 15 voids comprising the CAVITY parent sample, including the distribution of galaxies with respect to the void centres, as well as their colour and $r$-band absolute magnitude distributions. In Section~\ref{Analysis:local}, we further build upon the results of \cite{2026arXiv260617676A} to investigate the local environments of DR2 galaxies, providing a more complete view of both the local and large-scale environmental conditions in which these systems evolve.

\subsection{CAVITY voids characteristics}\label{Analysis:voids}
In Fig.~\ref{fig:Voids}, we show the distribution of VGs in three representative voids from the parent sample of CAVITY, namely voids 65, 405, and 941, for which some specific characteristics are discussed below. The left-hand panels show the distribution of void-centric distances, expressed as a fraction of $R_{\mathrm{e,void}}$, for all galaxies in each corresponding void. The right-hand panels show the corresponding colour–magnitude distributions for each void. The histograms on the top and right-hand sides display the distributions of $M_r $ and $g$-$r$, respectively, with photometric quantities taken from \cite{2012MNRAS.421..926P}. Similar figures for the remaining 12 voids in the parent sample are provided in Appendix~\ref{appendix:voids}.

Focusing on the spatial distribution of VGs across the 15 voids with respect to void centres, it becomes evident, as already reported in the \citetalias{2024A&A...691A.161G}, that the number of VGs increases with void-centric distance. In other words, the majority of VGs reside between $0.5$ and $0.8~R_{\mathrm{e,void}}$ (and beyond), leaving the central regions of voids relatively underpopulated. This trend is primarily a geometric effect, as spherical shells at larger void-centric distances encompass progressively larger volumes. Indeed, the cumulative number of galaxies approximately follows the expected $r^3$ dependence, indicating that the volume number density of galaxies remains broadly constant across the voids \citep{2012MNRAS.421..926P}. Nevertheless, noticeable void-to-void variations are present. For example, in some cases, the innermost regions ($\lesssim 0.2~R_{\mathrm{e,void}}$) still host a small number of galaxies, as seen in voids 65, 439, 622, and 921. In contrast, in void 405 the galaxy population remains sparse even at intermediate radii, with fewer than 10 galaxies detected within $0.3$–$0.5~R_{\mathrm{e,void}}$. In some systems, such as voids 921 and 941, a clear discontinuity is observed, with a small population of galaxies in the central regions and the majority concentrated at larger void-centric distances ($\gtrsim 0.5~R_{\mathrm{e,void}}$). While this feature may partly reflect the magnitude-limited nature of the SDSS sample, on which the CAVITY parent sample is based, which reduces the detectability of low-luminosity galaxies (see below), it may also indicate genuine differences in the spatial distribution of galaxies among individual voids \citep{2026A&A...710A.337T, 2026arXiv260617676A}.
\begin{figure}
\centering
    \includegraphics[width=0.85\columnwidth]{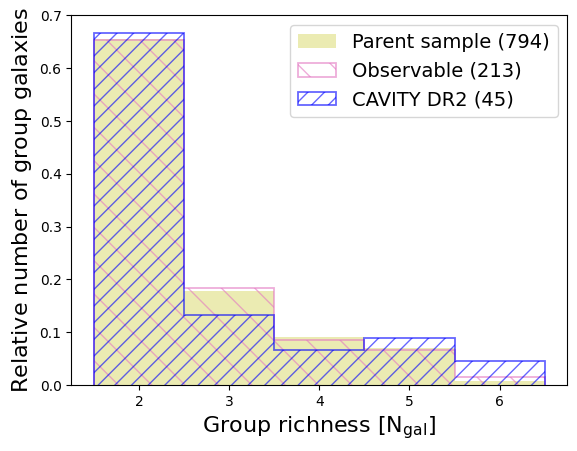} \\
    \includegraphics[width=0.85\columnwidth]{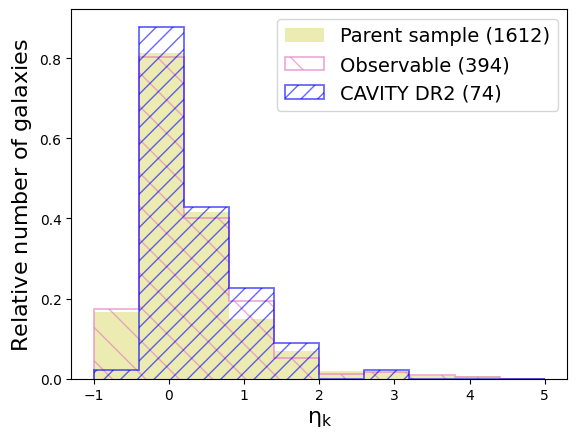}    
\caption{Distribution of the group richness ($N_{gal}$, upper panel) and projected local number density ($\rm \eta_k$, lower panel) for CAVITY galaxies. In the upper panel we only consider galaxies in groups with physically bound neighbours (i.e. neighbours within a projected distance d\,$\leq$\,450\,kpc and $\Delta v\,\leq\,\pm~160~\rm km\,s^{-1}$). In the lower panel we exclude galaxies with zero or one neighbour in their local environment (i.e. neighbours within d\,$\leq$\,1\,Mpc and $\Delta v\,\leq\,\pm~500~\rm km\,s^{-1}$). Yellow histograms denote galaxies in the CAVITY parent sample (794 and 1612 galaxies, respectively), while pink is used for the observable subsample (213 and 394, respectively) and blue for the final CAVITY DR2 sample (45 and 74 galaxies, respectively).}  \label{fig:sample_localenv}
\end{figure}

The voids in the CAVITY parent sample also differ significantly in their galaxy populations. In terms of hosted galaxies, voids 65, 139, and 487 are the most populated, each containing more than 200 VGs. At the opposite extreme, voids 622 and 941 host fewer than 40 galaxies and are among the least populated systems in the sample. It should be noted that the CAVITY parent sample was constructed by selecting only voids hosting at least 20 galaxies. Consequently, less populated systems present in the \citet{2012MNRAS.421..926P} catalogue are excluded, and the low-richness end of the void population is not fully sampled by CAVITY.

In general, the CAVITY parent sample is drawn from the SDSS DR7 spectroscopic sample, which is complete down to an $r$-band Petrosian magnitude of $r<$17.77 mag \citep{2002AJ....124.1810S}. This magnitude limit implies an increasing incompleteness towards the low-mass galaxies, particularly at higher redshift, and therefore, a lower observed abundance of dwarf galaxies within individual voids at higher distances is expected. For example, nearby voids 941, 482, and 738 ($z<$0.03) appear to host more low-luminosity systems, with many galaxies fainter than $M_r$ = -17.5 mag, whereas in more distant voids, such as 142 and 654 ($z>$0.04), the low-luminosity population appears largely absent. Consequently, the apparent scarcity of low-mass galaxies in such systems should not be interpreted as evidence for their intrinsic absence. Further deeper observations are required to properly constrain the faint-end VGs populations in these voids.

Nevertheless, even with this detection limit, notable differences are apparent in the luminosity distributions of VGs among different voids. In particular, completeness effects alone cannot fully account for the diversity observed among nearby voids. A notable example is provided by voids 941 and 738, located at comparable redshifts of $z\sim$ 0.013 and $z\sim$ 0.018, respectively. In both systems, galaxies fainter than $M_r $=-17 mag are detected, as expected given their proximity. Yet, while void 738 hosts galaxies spanning a wider luminosity range, the massive end of the distribution appears largely absent in void 941. Since both voids are subject to similar observational biases, this difference is unlikely to arise solely from completeness effects and instead may reflect intrinsic differences in their dynamical states or underlying physical properties.

In general, all 15 voids are predominantly populated by galaxies with blue photometric colours (i.e. $g$-$r$$<$0.6), consistent with results from previous studies \citep[e.g.][]{2004ApJ...617...50R, 2017A&A...600L...6K}. However, the fraction of red galaxies is not uniform across all the voids in the sample. While some voids, such as 487 and 142, exhibit a clear bimodality in $g$-$r$, indicative of a substantial red galaxy population, others, such as void 439, are overwhelmingly dominated by blue-cloud systems. The majority of red galaxies are also among the brightest, and therefore typically more massive, systems. However, some voids show a non-neligible population of low-luminosity red galaxies. A notable example is void 738, where a considerable fraction of galaxies with lower magnitudes display red colours. Part of this trend is related to the incompleteness affecting SDSS photometry, since the effective completeness limit at a given redshift depends on galaxies' average stellar population age and colour \citep[see Appendix A in][]{2026A&A...710A.337T}. Consequently, nearby voids provide a more complete census of low-mass red galaxies than their more distant counterparts.  

During the scheduling of CAVITY observations, two primary priorities were adopted when possible: (i) to obtain observations for at least 20 galaxies per void, where the number of observable galaxies exceeds this threshold, and (ii) to prioritise galaxies located in the inner regions of the voids, progressively extending towards larger void-centric radii. This strategy has been achieved for several voids, such as 355 and 738. However, in other systems, the target has not yet been fully reached, either because the central regions of some voids (e.g. voids 142 and 941) do not contain galaxies satisfying the CAVITY observational selection criteria or because of the limited number of observed galaxies as of May 2026 (e.g. void 65).

At the current stage of the survey, CAVITY has achieved coverage of more than 30\% of the observable galaxy population in nine voids (defined as the ratio of observed to observable galaxies within each void), and exceeds 70\% in three voids. Eight of the 15 voids are already sampled with at least 20 observed galaxies. The lowest completeness levels are found in voids 65 (9\%) and 487 (11\%), which are among the most populated systems in the parent sample. Future observations will continue to prioritise reaching the target of at least 20 observed galaxies per void, while also giving priority to the least well-sampled voids in order to improve the overall balance of the survey.

\subsection{Local environment characteristics}\label{Analysis:local}
The local environment of galaxies can be characterised through various approaches. Here, we adopt two widely used diagnostics for CAVITY VGs: group membership based on the work of \cite{2026arXiv260617676A}, tracing physically bound systems, and local galaxy density, measured through the abundance of neighbouring galaxies.

Starting from the \citet{2012MNRAS.421..926P} SDSS catalogue of voids, \cite{2026arXiv260617676A} used a Friend-of-Friend-like algorithm to find group galaxies in voids with $z$<0.08, containing 24931 VGs. Each void is evaluated independently, considering only the galaxies that reside in it. 
First, they searched for neighbour galaxies with line-of-sight velocity difference $\Delta v\,\leq\,\pm~500~\rm km\,s^{-1}$ within 1\,Mpc in projected distance \citep{2015A&A...578A.110A}. If a galaxy has no neighbours within that volume (59\% of VGs), these are classified as singlets (i.e. galaxies that are not physically bound to any other detected galaxy). Conversely, they are identified as physically bound groups if the galaxies have neighbours separated by a projected distance d\,$\leq$\,450\,kpc with $\Delta v\,\leq\,\pm~160~\rm km\,s^{-1}$. Galaxies that have neighbours within the defined volume but are not physically bound with them; hereafter 'loose' galaxies, act as tracers of the substructure inside the voids (29\% of the galaxies in the voids). Following this methodology, they identified a sample of 1367 physically bound groups, composed of 3040 galaxies (12\% of the VGs), where the most massive galaxy in each group is defined as the central one. 

When translating these numbers to the CAVITY parent sample (15 voids, 4866 galaxies), 42\% of the galaxies are singlets (2055 galaxies), 16\% are in groups with $\rm N_{gal}\,\ge\,2$ (794 galaxies), and 42\% are loose galaxies (2017 galaxies). In comparison, 213 group galaxies (19\%) are included in the observable sample and 45 galaxies (22\%) in the DR2 sample. The rest of the galaxies are singlets (427 galaxies in the observable sample and 62 in the DR2 sample), and loose galaxies (475 galaxies in the observable sample and 93 in the DR2 sample). Table~\ref{tab:grichness} summarises the number of group, singlets, and loose galaxies in each sample. 
Most of the group galaxies belong to groups with two physically bound galaxies. This is evident in the upper panel of Fig.~\ref{fig:sample_localenv}, where we present the distribution of group richness (number of physically bound neighbours in each group) for group galaxies in the parent, the observable, and the CAVITY DR2 samples.

Overall, the fraction of galaxies belonging to different group categories in CAVITY DR2 closely follows that of the CAVITY parent sample across the 15 surveyed voids. Furthermore, the DR2 sample spans the full range of group richness present in the CAVITY parent sample, including its richest groups. This makes DR2 a well-suited sample for investigating the role of local environment in shaping the evolution of VGs.

Another common method to quantify the local environment of VGs is to use the local galaxy number density (i.e. the number of galaxies per unit volume) within a fixed field radius. We estimated the galaxy number density, $\eta_k$, as: 
\begin{equation}
    \eta_k = \log \left(\frac{N_{neigh} - 1}{\frac{4}{3}\pi d^3} \right), 
\end{equation}
where $N_{neigh}$ is the number of local neighbours and $d$ is the projected distance to the k$^{th}$ nearest neighbour, with $k=5$ or less if there are not enough neighbours in the field \citep{1985ApJ...298...80C, 2007A&A...472..121V,2015A&A...578A.110A}. The lower the $\eta_k$ value, the less dense the local environment is. To compare this quantification of the local environment with the group environment, we look for local neighbours within the same void, with a projected field radius of 1\,Mpc, and with a line-of-sight velocity difference of $\Delta v$\,=\,500\,km\,s$^{-1}$. The distribution of $\eta_k$ for galaxies in the CAVITY DR2 sample, in comparison to the parent sample and observable galaxies, is shown in the lower panel of Fig.~\ref{fig:sample_localenv}. Note that by definition, the $\eta_k$ can not be computed if a galaxy has one or no neighbours. This happens for the 67\% of the galaxies in the parent sample, with similar numbers for the observable and DR2 samples (65\% and 63\% respectively), which confirms the extremely low density environment within voids.

Despite the absence of any selection criterion based on local galaxy density, both the observable and DR2 samples provide a representative view of the ($\eta_k$) distribution of the CAVITY parent sample. A slight underrepresentation is observed only for galaxies with (-1 $<$ $\eta_k$ $<$ 0), corresponding to systems with two to three neighbours within the defined volume.

\begin{table}
    \centering
    \caption{\label{tab:grichness}Numbers of group, singlets, and loose galaxies.}
    \begin{tabular}{cccc}
    \hline\hline
    Sample     & group        & singlets    & loose         \\ 
    \hline
    Parent     & 794 (16\%)   & 2055 (42\%) & 2017 (42\%)   \\
    Observable & 213 (19\%)   &  427 (38\%) &  475 (43\%)   \\ 
    DR2        &  45 (22.5\%) &   62 (31\%) &   93 (46.5\%) \\     
    \hline
     \end{tabular}
    \tablefoot{Numbers of group, singlets, and loose galaxies in the parent, observable, and DR samples, and the corresponding percentages with respect to the total number of VGs in each sample.}
\end{table}


\section{Summary and conclusions}\label{Summary}
In this article, we present the second data release (DR2) of the Calar Alto Void Integral-field Treasury Survey (CAVITY), including a detailed description of the sample selection, observing strategy, data reduction, and quality-control procedures. This release comprises 200 integral-field spectroscopy (IFS) data cubes, including the 100 galaxies previously made available in DR1 and 100 newly released science-grade, quality-validated galaxies. All data were obtained with the PMAS/PPAK spectrograph at the 3.5-m Calar Alto Observatory (CAHA) using the V500 set-up, which covers the optical wavelength range 3745–7500 \AA\ at a spectral resolution of 6.0 \AA\ (FWHM). The data were reduced using version 1.2 of the official, automated CAVITY reduction pipeline \citepalias[see][]{2024A&A...691A.161G}. {The main improvement provided by DR2 is the increase in statistical power. By doubling the number of galaxies released in DR1, while retaining the same selection criteria and observing strategy, DR2 provides a larger and homogeneous sample for statistical studies. This expanded sample will allow more robust constraints on correlations between galaxy properties and their large-scale environment, and will provide a stronger basis for subsequent studies of galaxy evolution within the CAVITY project.}

Overall, the CAVITY DR2 sample closely follows the distribution of the CAVITY parent sample across the 15 selected voids, spanning a broad range of stellar masses, sizes, morphologies, star formation activity, and ionisation mechanisms. Compared to DR1, the current release reduces the slight overrepresentation of bright galaxies ($M_r \sim -20$ mag) by including a larger fraction of fainter systems. Nevertheless, the low-mass, low-luminosity regime ($M_r > -18$ mag) remains largely underrepresented due to the limited spatial sampling and observational constraints imposed by the current instrumental setup for faint and compact galaxies. As of March 2026, CAVITY has achieved a coverage exceeding 30\% of the observable galaxy population in nine voids and more than 70\% in three of them. Overall, the 15 voids comprising the CAVITY parent sample span a broad range of properties, including different sizes, galaxy richness, and galaxy populations with diverse stellar masses and star-formation states. Due to the magnitude-limited nature of the SDSS survey from which the parent sample is drawn, more distant voids are preferentially populated by brighter and more massive galaxies, while the detection of low-mass dwarf systems becomes increasingly incomplete. To characterise the local environment of VGs, we adopted two complementary approaches: galaxy group membership and local galaxy density. Although the DR2 sample was assembled without explicitly selecting galaxies to reproduce the distribution of local environmental properties, it nevertheless provides a representative view of the local environments found in the CAVITY parent sample, spanning the full range of group richness.

With the second data release, CAVITY has now delivered a substantial fraction of its final legacy sample, providing public access to approximately two-thirds of the expected survey data. Observations have already been completed for more than 300 VGs, and the remaining data will continue to be processed through updated versions of the reduction pipeline and quality-control procedures. The final CAVITY data release will provide an homogeneous sample of more than 300 science-grade IFS data cubes of VGs.
In addition to increasing the sample size, the next stages of CAVITY will focus on improving the coverage of individual voids, aiming to observe at least 20 galaxies per void whenever possible and to provide a more complete mapping of the radial distribution of galaxies within each void. Particular emphasis will be placed on currently under-sampled voids to ensure a more homogeneous representation of the diverse void population.

\begin{acknowledgements}
{We thank the referee for the relevant suggestions that
have helped us to improve the final version. }This paper is (partially) based on observations collected at the Centro Astronómico Hispano en Andalucía (CAHA) at Calar Alto, operated jointly by Junta de Andalucía and Consejo Superior de Investigaciones Científicas (IAA-CSIC), under the CAVITY legacy project. The CAVITY project acknowledges financial support by the research projects AYA2017-84897-P, PID2020-113689GB-I00, PID2020-114414GB-I00, and PID2023-149578NB-I00 funded by the Spanish Ministry of Science and Innovation (MCIN/AEI/10.13039/501100011033) and by FEDER/UE; the project A-FQM-510-UGR20 funded by FEDER/Junta de Andalucía-Consejería de Transformación Económica, Industria, Conocimiento y Universidades/Proyecto; by the grants P20-00334 and FQM108, funded by Junta de Andalucía; and by Consejería de Universidad, Investigación e Innovación (Junta de Andalucía) and Gobierno de España and European Union NextGenerationEU through grant AST22\_4.4. R.G.D., A.C., and R.G.B. acknowledge financial support from the Severo Ochoa grant CEX2021-001131-S funded by MCIN/AEI/ 10.13039/501100011033, and PID2022-141755NB-I00. MAF and PVB acknowledge support from the Emergia program (EMERGIA20\_38888) from Consejer\'ia de Universidad, Investigaci\'on e Innovaci\'on de la Junta de Andaluc\'ia. SBD acknowledges financial support from the grant AST22.4.4, funded by Consejería de Universidad, Investigación e Innovación and Gobierno de España and Unión Europea – NextGenerationEU, and project PID2021-122544NB-C43. YGK acknowledges financial support from PREP2023-001684 funded by MCIU/AEI/10.13039/501100011033 and the FSE+. TRL acknowledges support from Ramón y Cajal fellowship (RYC2023-043063-I, financed by MCIU/AEI/10.13039/501100011033 and by the FSE+). GTR acknowledges financial support from the research project PRE2021-098736, funded by MCIN/AEI/10.13039/501100011033 and the FSE+.  HMC acknowledges support from the Institut Universitaire de France. IMC acknowledges funding from Agencia Nacional de Investigación y Desarrollo / Fondo ALMA 2025 / 31250023. J.F-B acknowledges support from the PID2022-140869NB-I00 grant from the Spanish Ministry of Science and Innovation. AFM acknowledges support from RYC2021-031099-I and PID2024-162088NB-I00 of MICIN/AEI.  SS acknowledges the support from the Alexander von Humboldt Foundation.  L.G. acknowledges financial support from MCIN and AEI
10.13039/501100011033 under projects PID2023-151307NB-I00,
CEX2020-001058-M, and by the MaX-CSIC Excellence Award MaX4-SOMMA-ICE. SFS acknowledges the support by CBF-2025-I-236 project granted by the Secretaría de Ciencia, Humanidades, Tecnología e Innovación (SECIHTI) of
the Mexican Federal Government, and the PID2022-136598NB-C31 (ESTALLIDOS) grant by the Spanish Ministery of Science and Innovation (MCINN).

      Funding for SDSS-III has been provided by the Alfred P. Sloan Foundation, the Participating Institutions, the National Science Foundation, and the U.S. Department of Energy Office of Science. SDSS-III is managed by the Astrophysical Research Consortium for the Participating Institutions of the SDSS-III Collaboration including the University of Arizona, the Brazilian Participation Group, Brookhaven National Laboratory, Carnegie Mellon University, University of Florida, the French Participation Group, the German Participation Group, Harvard University, the Instituto de Astrofísica de Canarias, the Michigan State/Notre Dame/ JINA Participation Group, Johns Hopkins University, Lawrence Berkeley National Laboratory, Max Planck Institute for Astrophysics, Max Planck Institute for Extraterrestrial Physics, New Mexico State University, New York University, Ohio State University, Pennsylvania State University, University of Portsmouth, Princeton University, the Spanish Participation Group, University of Tokyo, University of Utah, Vanderbilt University, University of Virginia, University of Washington, and Yale University. The SDSS-III web site is {\tt \href{http://www.sdss3.org/}{http://www.sdss3.org}}.
      We are also grateful for the computing resources and related technical support provided by PROTEUS, the supercomputing center of Institute Carlos I in Granada, Spain.
      This research made use of \cite{python} v3.11 programming language.
      This work made use of Astropy (\url{http://www.astropy.org}) a community-
      developed core Python package and an ecosystem of tools and resources for astronomy \citep{2013A&A...558A..33A, 2018AJ....156..123A, 2022ApJ...935..167A}; SciPy \citep{2020SciPy-NMeth}; Numpy \citep{2011CSE....13b..22V, harris2020array}; Pandas \citep{mckinney-proc-scipy-2010, 2022zndo...3509134T}; and Matplotlib \citep{2007CSE.....9...90H}.
      
\end{acknowledgements}

\bibliographystyle{aa} 
\bibliography{main}

\begin{appendix}

\section{Access to the CAVITY DR2 data}\label{appendix:Data_access}

All CAVITY DR2 data cubes and catalogues are publicly available through the survey's  webpage\footnote{\url{https://cavity.caha.es/}}, which also provides project updates and information on future CAVITY and CAVITY+ data releases. Moreover, the dedicated DR2 webpage\footnote{\url{https://cavity.caha.es/data/dr2/}} provides access to the released data products such as catalogues and PPAK images. The database structure and data tables remain unchanged from the first data release, and are described in detail in the \citetalias{2024A&A...691A.161G} paper.

The CAVITY database is built using the open-source Daiquiri framework \citep{2020ASPC..527..395G}, which provides functionalities such as news updates, user comments, and an SQL query interface. The CAVITY catalogues are distributed in the commonly used CSV and FITS formats and are fully compatible with the Virtual Observatory (VO) standards. The catalogues can be accessed through the SQL query interface available on the CAVITY query webpage\footnote{\url{https://cavity.caha.es/query/}} as well as through the Table Access Protocol \citep[TAP;][]{2011ASPC..442..603D}. Using SQL or Astronomical Data Query Language (ADQL; the IVOA standard query language) queries, users can access the complete DR2 dataset and other CAVITY catalogues, with the possibility of applying customised selection criteria tailored to specific scientific goals. To demonstrate the capabilities of the database and facilitate its use, an example query retrieving PPAK data cubes, images, and selected columns from the master catalogue for two galaxies (with survey IDs 34100 and 34101) in the DR2 sample is provided below. Additional examples for all currently available CAVITY data releases can be found in the examples section of the query webpage.

\begin{lstlisting}[language=SQL]
SELECT m.galaxy, m.ra, m.dec, m.redshift,
       m.M_star, c.file, 
       i.img_footprint, i.img_moment
FROM dr2.Master as m, dr2.PPAK_images as i, 
     dr2.PPAK_cubes as c
WHERE m.galaxy = i.galaxy 
  AND m.galaxy = c.galaxy 
  AND m.galaxy IN (34100,34101)
\end{lstlisting}
\section{Distribution of CAVITY DR2 galaxies in voids}\label{appendix:voids}
In Figs.~\ref{fig:VoidsI}, \ref{fig:VoidsII}, and \ref{fig:VoidsIII}, we present the distribution of VGs in the CAVITY parent sample (excluding  voids 65, 405, and 941, whose distributions  are presented  in the main text in Section~\ref{Analysis:voids}).

\begin{figure*}
\centering
    \includegraphics[width=0.8\textwidth]{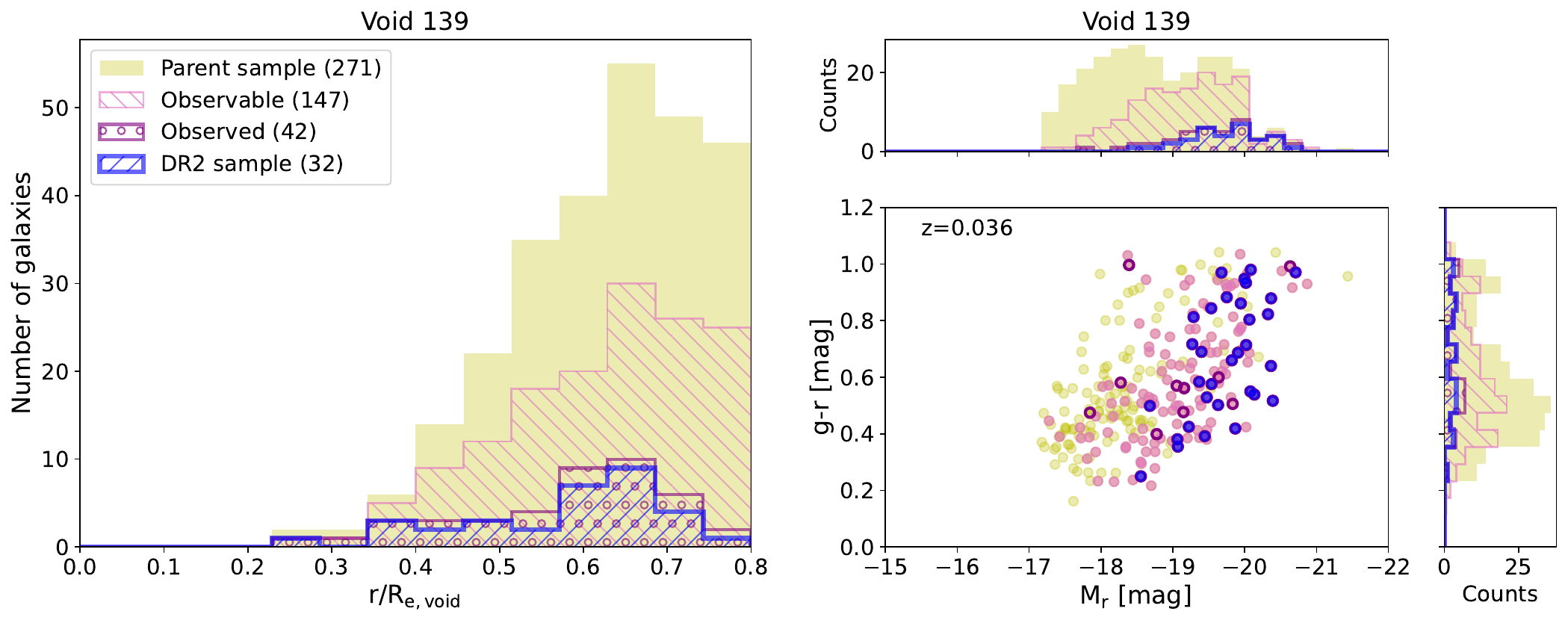}
    \includegraphics[width=0.8\textwidth]{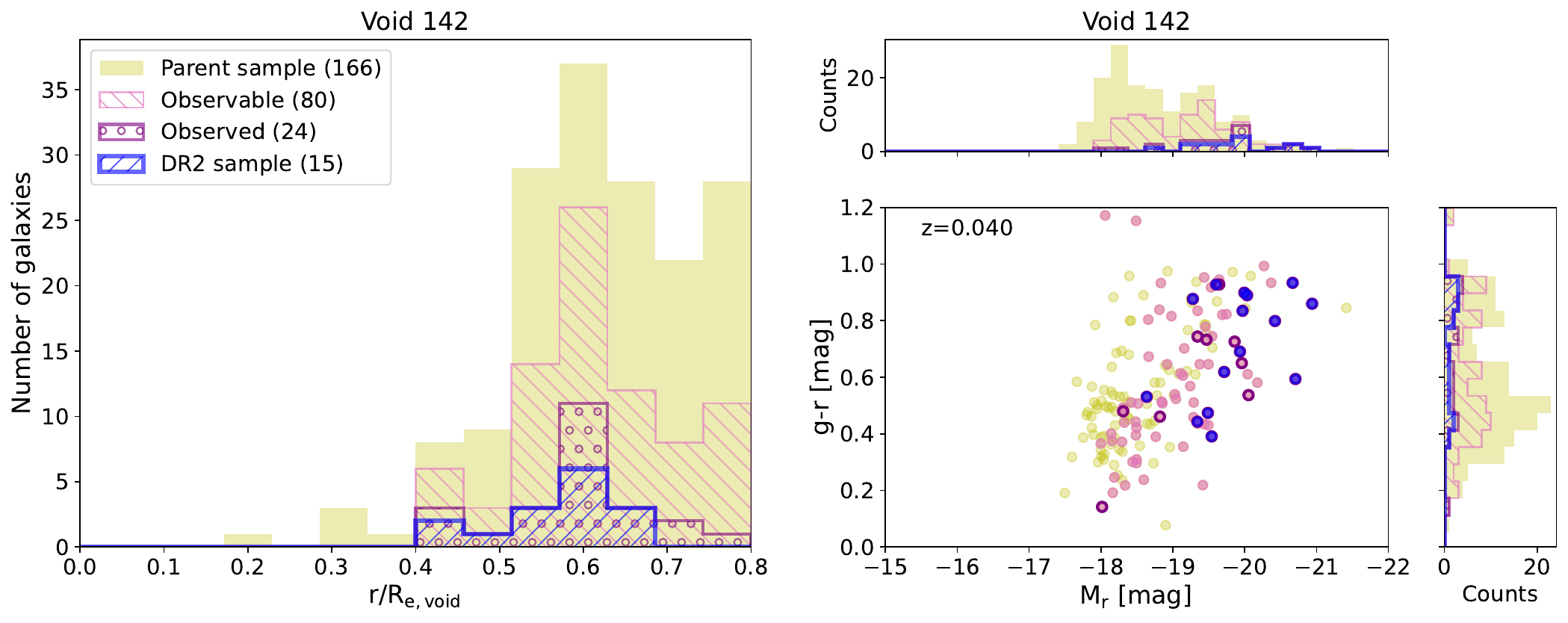}
    \includegraphics[width=0.8\textwidth]{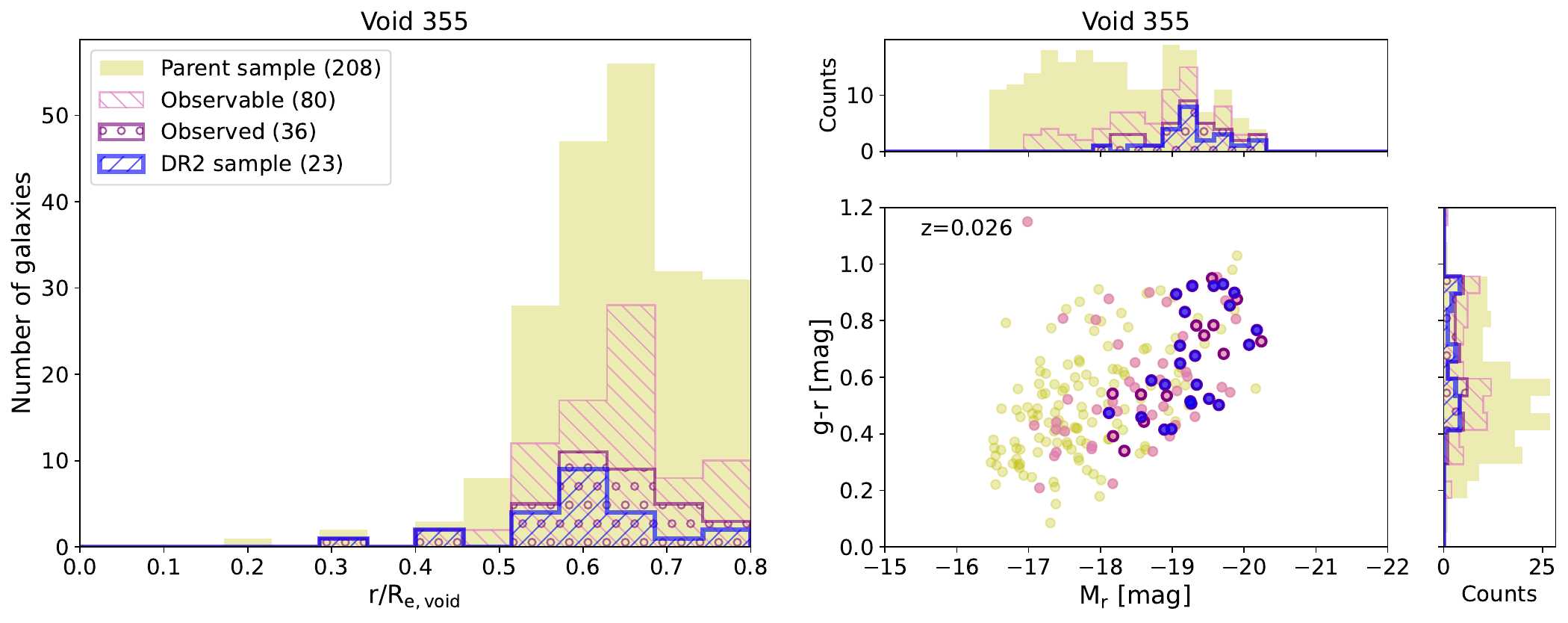}
    \includegraphics[width=0.8\textwidth]{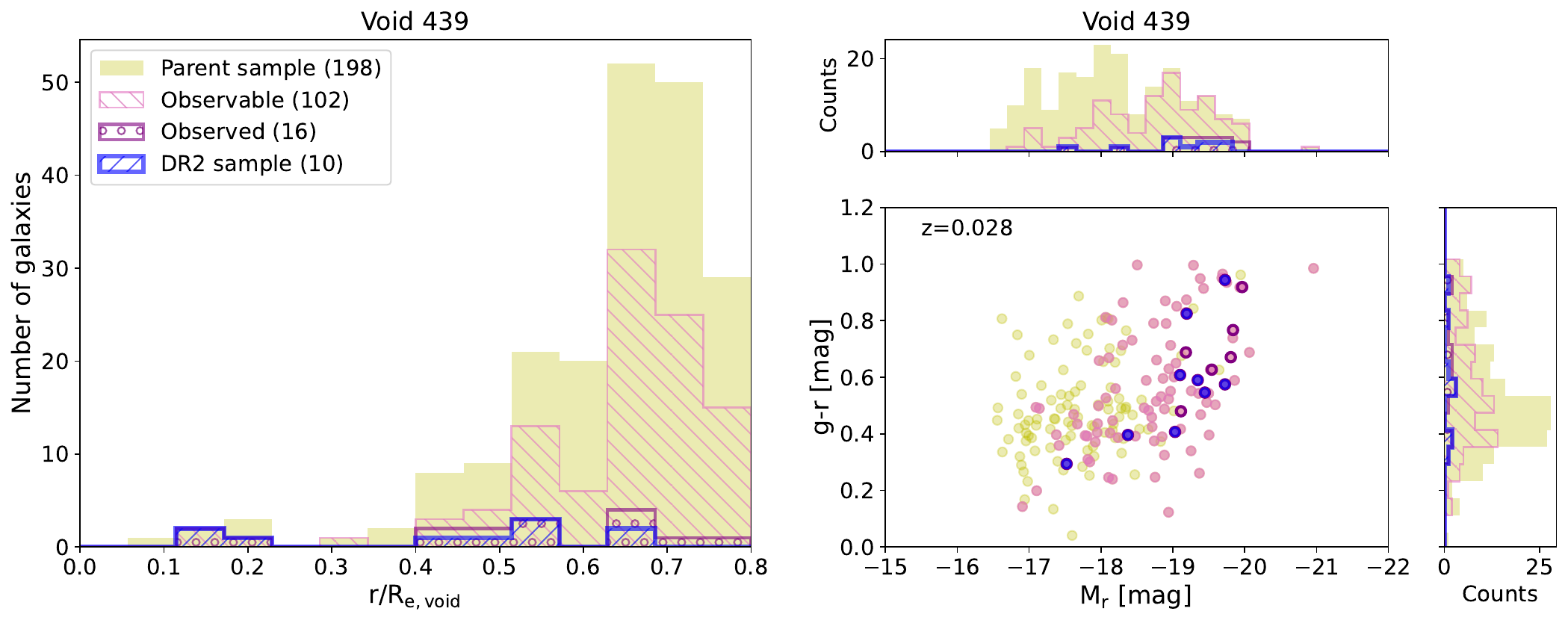}
\caption{Similar to Fig.~\ref{fig:Voids}, shown here for the remaining voids in the CAVITY parent sample.}  \label{fig:VoidsI}
\end{figure*}

\begin{figure*}
\centering
    \includegraphics[width=0.8\textwidth]{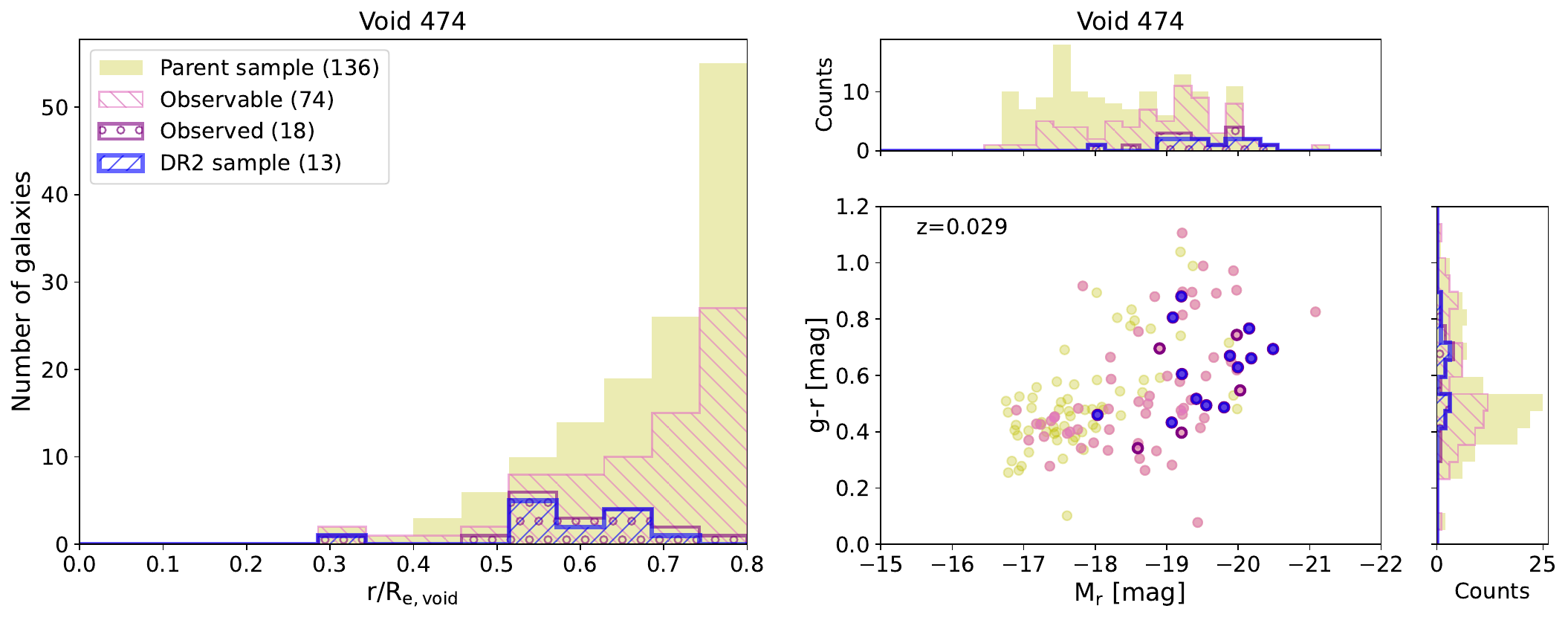}
    \includegraphics[width=0.8\textwidth]{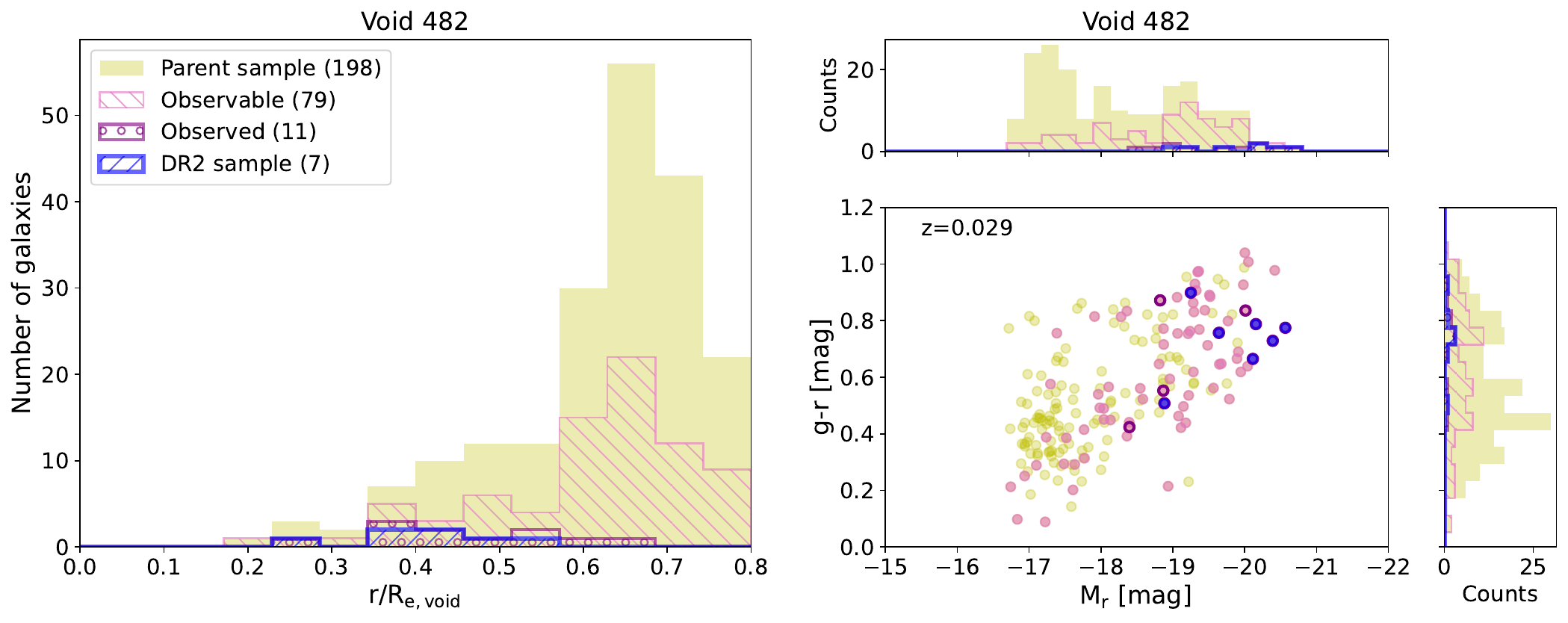}
    \includegraphics[width=0.8\textwidth]{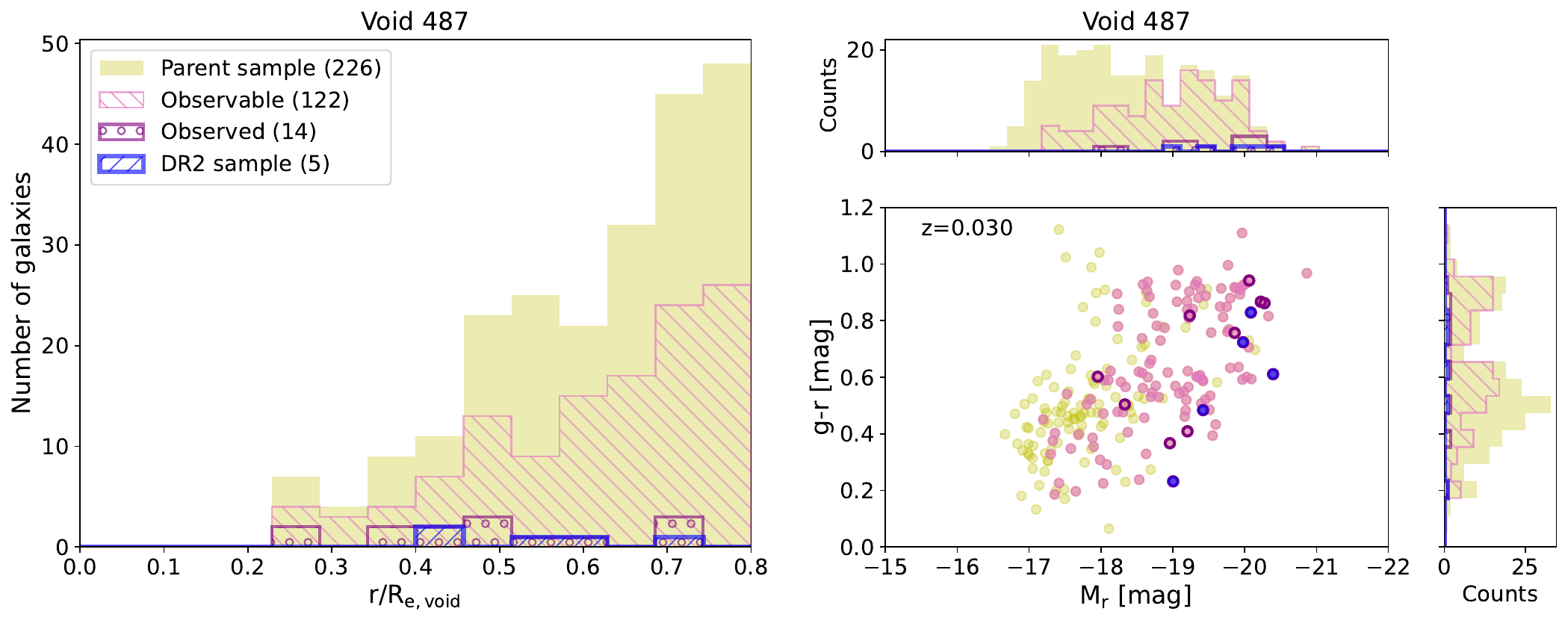}
    \includegraphics[width=0.8\textwidth]{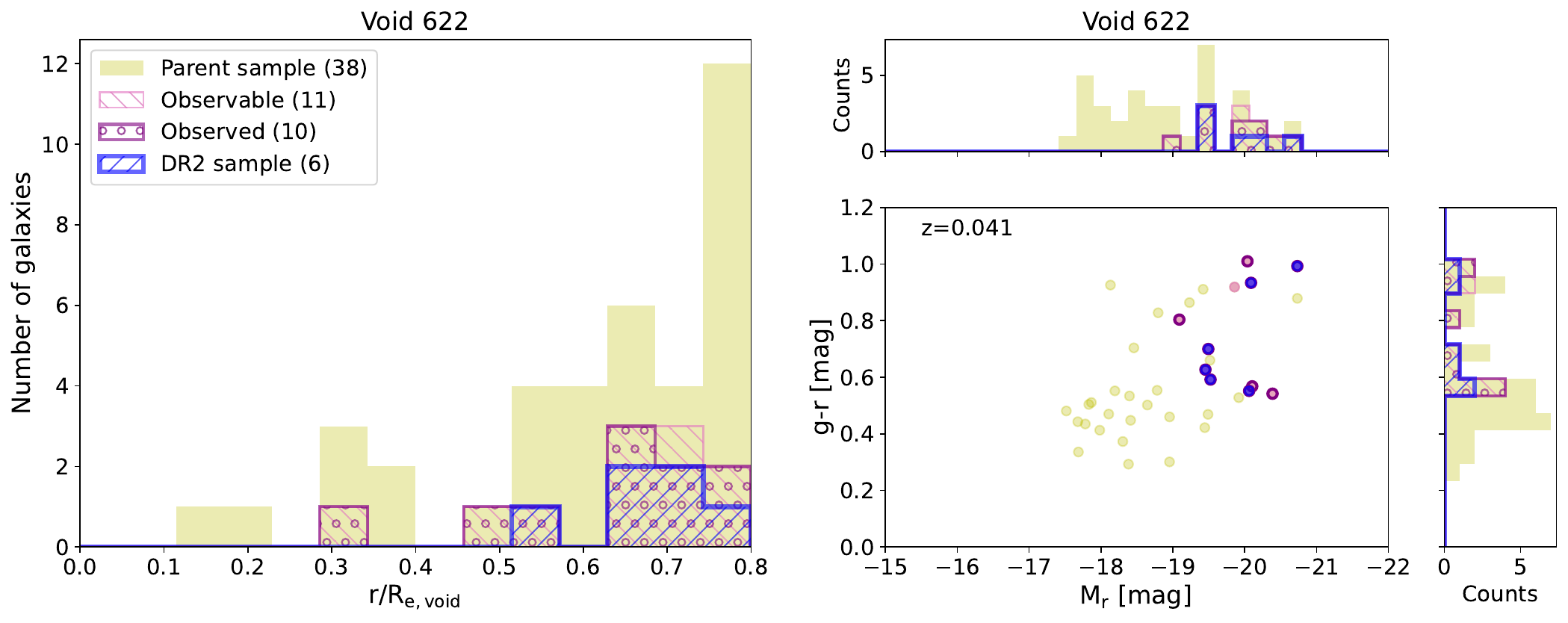}
\caption{Continued.}  \label{fig:VoidsII}
\end{figure*}
\begin{figure*}
\centering
    \includegraphics[width=0.8\textwidth]{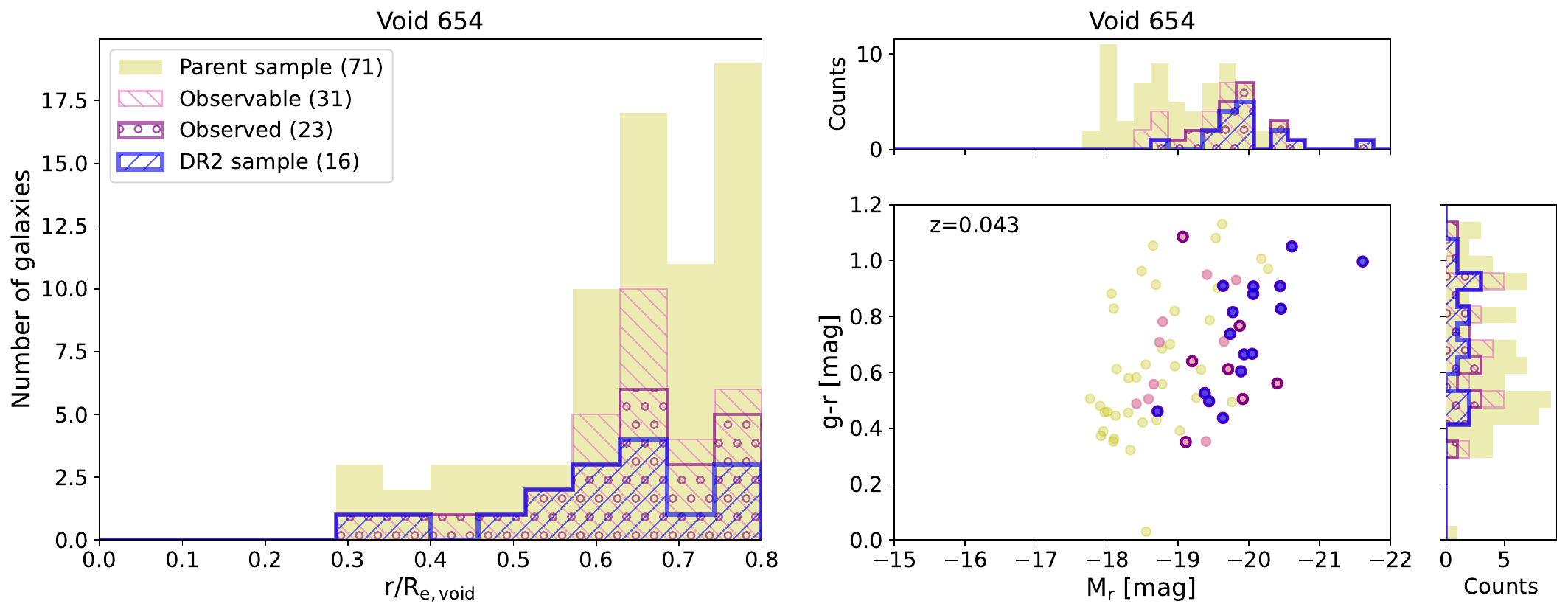}
    \includegraphics[width=0.8\textwidth]{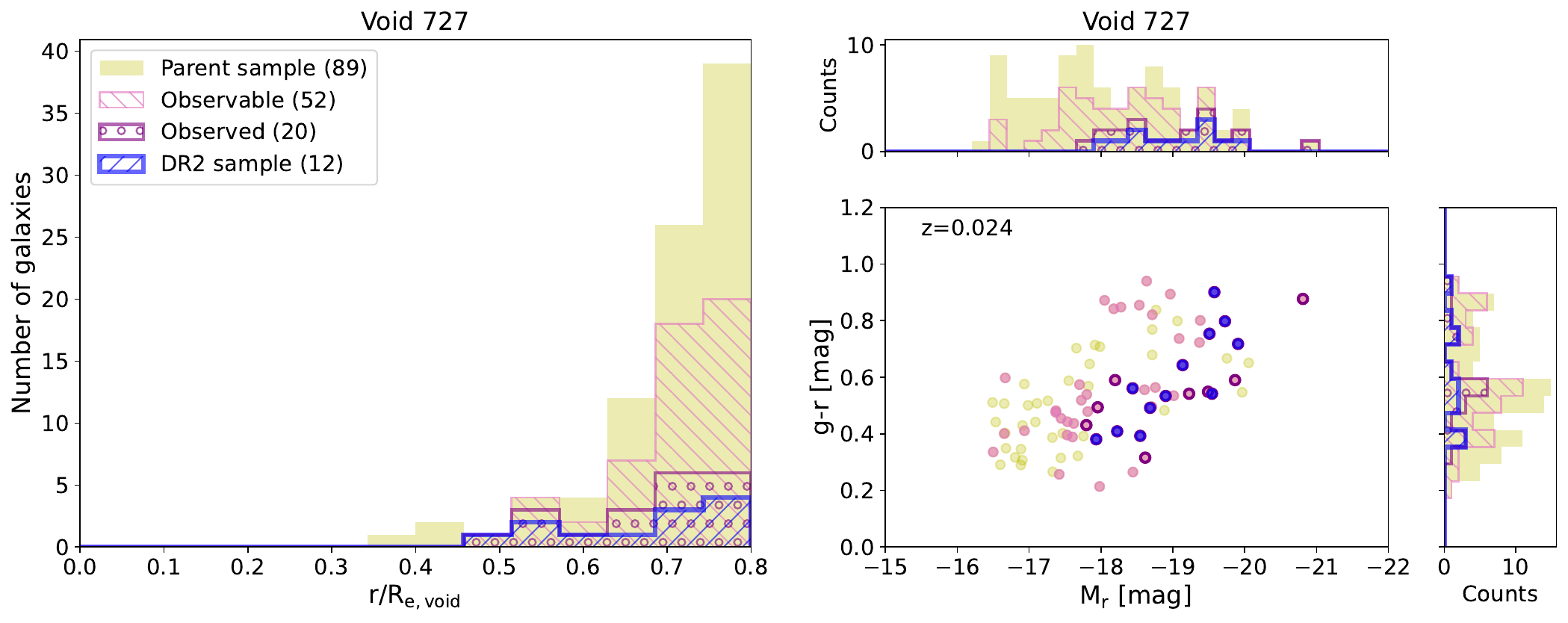}
    \includegraphics[width=0.8\textwidth]{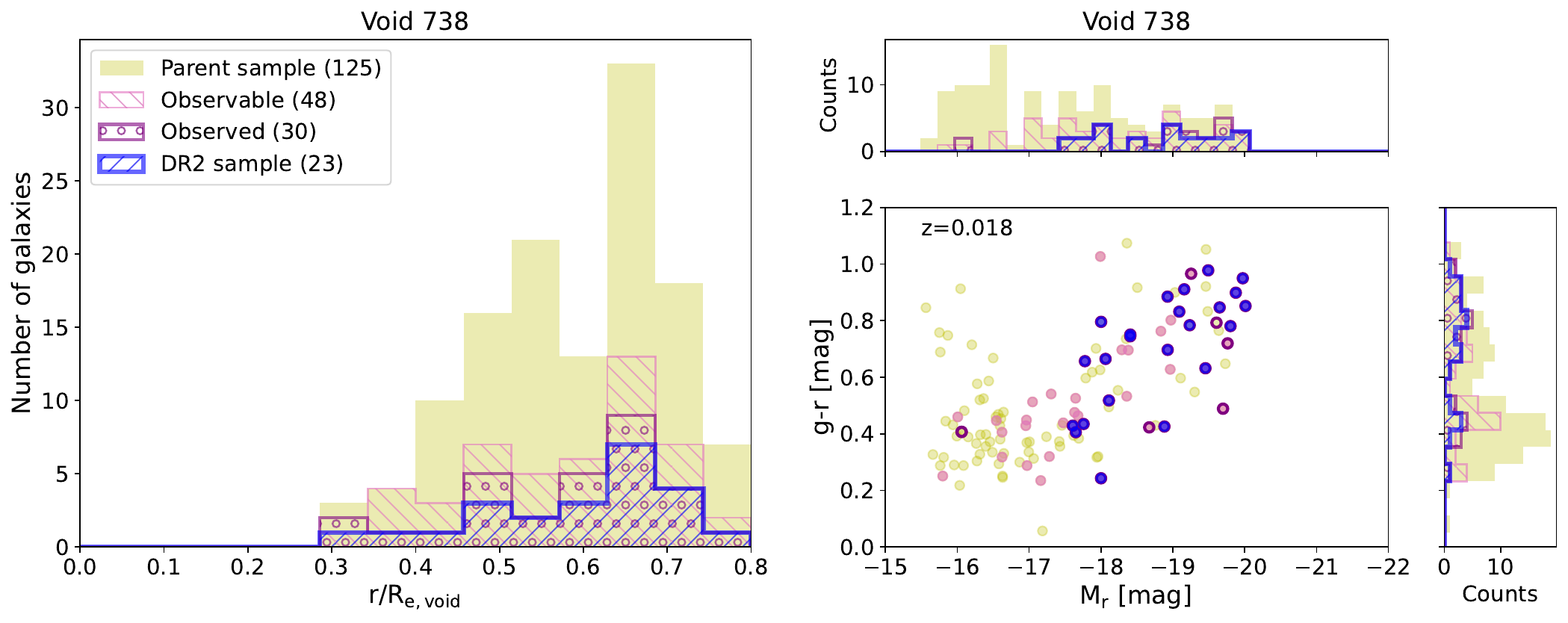}
    \includegraphics[width=0.8\textwidth]{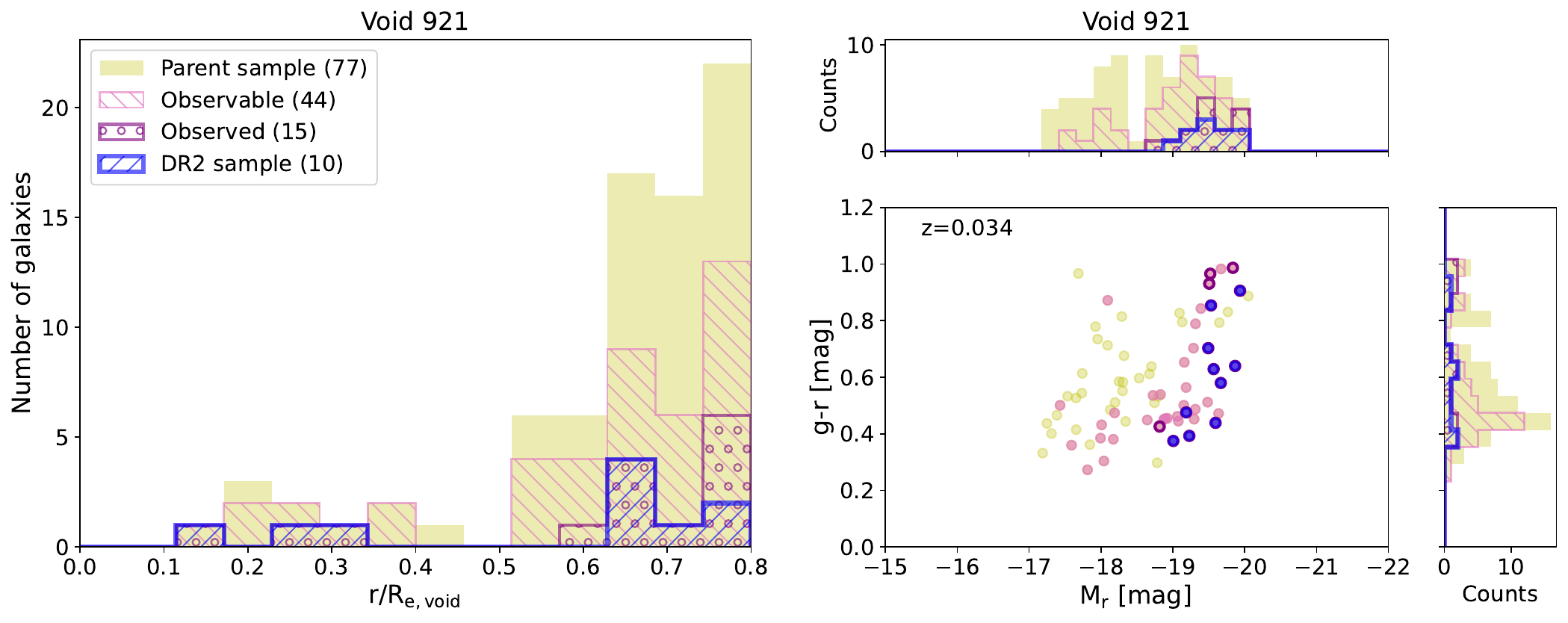}
\caption{Continued.}  \label{fig:VoidsIII}
\end{figure*}

\end{appendix}

\end{document}